\documentclass[iop]{emulateapj}
\slugcomment{{Accepted for publication in the Astrophysical Journal}}
\usepackage{epsfig, natbib, graphicx}
\def\arcsec{\hbox{$^{\prime\prime}$}}

\begin{document}

\title{\textit{Chandra} and \textit{XMM-Newton} studies of the supernova remnant G292.2--0.5 associated with the pulsar J1119--6127}
\author{Harsha S. Kumar \altaffilmark{1}, Samar Safi-Harb\altaffilmark{1,2} and Marjorie E. Gonzalez\altaffilmark{3}}
\altaffiltext{1}{Department of Physics \& Astronomy, University of Manitoba, Winnipeg, MB, R3T 2N2, Canada; harsha@physics.umanitoba.ca.}
\altaffiltext{2}{Canada Research Chair; samar@physics.umanitoba.ca.}
\altaffiltext{3}{Department of Physics \& Astronomy, University of British Columbia, Vancouver, BC, V6T 1Z1, Canada;
gonzalez@phas.ubc.ca.}

\begin{abstract}

We present the first detailed imaging and spatially resolved spectroscopic study of the Galactic supernova remnant (SNR) G292.2--0.5, associated with the high-magnetic field radio pulsar (PSR) J1119--6127, using \textit{Chandra} and \textit{XMM-Newton}. The high-resolution X-ray images reveal a partially limb-brightened morphology in the west, with diffuse emission concentrated towards the interior of the remnant unlike the complete shell-like morphology observed at radio wavelengths. The spectra of most of the diffuse emission regions within the remnant are best described by a two-component thermal+non-thermal model. The thermal component is described by a plane-parallel, non-equilibrium ionization plasma model with a temperature $kT$ ranging from 1.3$^{+0.3}_{-0.2}$ keV in the western side of the remnant to 2.3$^{+2.9}_{-0.5}$ keV in the east,  a column density increasing from 1.0$^{+0.1}_{-0.6}$$\times$10$^{22}$~cm$^{-2}$ in the west to 1.8$^{+0.2}_{-0.4}$$\times$10$^{22}$~cm$^{-2}$ in the east, and a low ionization timescale ranging from  (5.7$^{+0.8}_{-0.7}$)$\times$10$^9$ cm$^{-3}$~s  in the SNR interior to (3.6$^{+0.7}_{-0.6}$)$\times$10$^{10}$~cm$^{-3}$~s in the western side -- suggestive of expansion of a young remnant in a low-density medium. 
The spatial and spectral differences across the SNR are consistent with the presence of a dark cloud in the eastern part of the SNR, absorbing the soft X-ray emission, as also revealed by the optical image of that region. The spectra from some of the regions also show slightly enhanced metal abundances from Ne, Mg and Si, hinting at the first evidence for ejecta heated by the reverse shock. Comparing our inferred metal abundances to core-collapse nucleosynthesis models yields, we estimate a high progenitor mass of $\sim$30$M_{\sun}$ suggesting a type Ib/c supernova.
We confirm the presence of non-thermal X-ray emission from regions close to the pulsar, with the emission characterized by a power-law model with a hard photon index similar to that seen in the compact pulsar wind nebula. We estimate an SNR age range between 4.2~kyr (free expansion phase) and 7.1~kyr (Sedov phase) at an assumed distance of 8.4~kpc,  a factor of a few higher than the measured pulsar's age upper limit of 1.9~kyr.

\end{abstract}


\section{Introduction}

The X-ray study of supernova remnants (SNRs) using satellites like \textit{Chandra} and \textit{XMM-Newton} with unprecedented sensitivity and spatial/spectral resolution has revolutionized our understanding of SNRs structure and dynamics. The X-ray emission from SNRs provides a wealth of information on the remnant properties such as age, energetics, temperature, and morphology, in addition to probing the composition of the supernova ejecta and the interaction of the SNR with the ambient interstellar medium (ISM). 

G292.2--0.5 is a SNR associated with the high-magnetic field radio pulsar (PSR) J1119--6127, which was discovered in the Parkes multibeam pulsar survey (Camilo et al. 2000).  PSR J1119--6127 has a relatively large rotation period $P$=408 ms and period derivative $\dot{P}$=4$\times$10$^{-12}$ s~s$^{-1}$, a characteristic age $\tau_c$=$P$/2$\dot{P}$=1.6~kyr, and a spin-down luminosity of $\dot{E}$=2.3$\times$10$^{36}$ ergs~s$^{-1}$. The pulsar's measured braking index, $n$=2.9 (Camilo et al. 2000), has been recently refined using more than 12 years of radio timing data yielding an $n$ value of  2.684$\pm$0.002 (Weltevrede et al. 2011), implying an upper limit on its age of 1.9~kyr. Its surface magnetic field strength\footnote{The pulsar's dipole magnetic field estimated as $B$ (Gauss) $\approx$3.2$\times$10$^{19}$($P/\dot{P}$)$^{1/2}$ where $P$ is in seconds.} is $B$$\sim$4.1$\times$10$^{13}$~G, making J1119--6127 a high-magnetic field radio pulsar with $B$ at the limit between the ``classical" rotation-powered pulsars and magnetars.  Radio observations made with the \textit{Australia Telescope Compact Array} (ATCA) revealed a non-thermal shell of $\sim$15$\arcmin$ in diameter and which has been identified as the SNR G292.2--0.5 (Crawford et al. 2001). Using the \textit{Advanced Satellite for Cosmology and Astrophysics} (ASCA) and the \textit{R\"oentgensatellit} (ROSAT) satellites, Pivovaroff et al. 2001 (hereafter P01) detected a hard point-like source $\sim$1$\arcmin$.5 from the position of the radio pulsar.  Also, an extended X-ray emission was observed from a circular region of diameter $\sim$17$\arcmin$ coincident with the SNR shell detected at radio wavelengths. The X-ray emission from the remnant was fitted using either thermal models of temperature $kT$$\sim$4 keV or a non-thermal power-law (PL) model of photon index $\Gamma$$\sim$2 (P01).  However, due to the limited data, these models could not fully characterize the spectral nature of the remnant (P01).

A \textit{Chandra} observation of PSR~J1119--6127, initially carried out in 2002, detected the X-ray counterpart to the radio pulsar and
resolved for the first time a 3$\arcsec$$\times$6$\arcsec$ faint pulsar wind nebula (PWN) associated with it (Gonzalez \& Safi-Harb 2003). The high-resolution \textit{Chandra} data also made possible an imaging and spatially resolved spectroscopic study of the western side of the SNR which decoupled the emission of the remnant from the pulsar and other point sources in the field (Gonzalez \& Safi-Harb 2005; hereafter GSH05). GSH05 interpreted the diffuse emission from the western side of the remnant as due to non-thermal emission characterized by a PL model with a photon index $\Gamma$$\sim$1--2 originating from the interior, plus thermal emission from the outer regions characterized by a non-equilibrium ionization model with a temperature $kT$$\sim$0.9--2.4~keV, an ionization timescale $n_et$$\sim$(3.3--12)$\times$10$^{10}$~cm$^{-3}$~s, and with sub-solar abundances indicating emission from shocked ISM. The SNR/PSR system observed with \textit{XMM-Newton} in 2003 detected unusually high pulsed fraction (74\%$\pm$14\%) associated with the pulsar in the 0.5--2.0 keV energy range (Gonzalez et al. 2005). 

A new and deep \textit{Chandra} observation of the eastern side of the remnant was undertaken in 2004 to complement the study of the western side of the remnant by GSH05. The combined analysis of the data confirmed the detection of the compact PWN associated with the pulsar and revealed a long southern jet ($\sim$6$\arcsec$$\times$21$\arcsec$) emanating from the position of the pulsar (Safi-Harb \& Kumar 2008).  This observation also allowed for resolving the pulsar's spectrum from its associated nebula, and as a result pinning down its X-ray spectrum in comparison to other high-magnetic field radio pulsars (Safi-Harb \& Kumar 2008). The kinematic distance to the SNR/pulsar system was estimated to be 8.4$\pm$0.4 kpc using HI absorption measurements (Caswell et al. 2004). As well, the SNR boundary was estimated to be an ellipse with the major and minor axes at 18$\arcmin$ and 17$\arcmin$, respectively, thus approximating to a diameter of $\sim$17$\arcmin$.5. In this paper, we adopt Caswell et al. (2004) values for the distance $D_{8.4}$=$D$/8.4 kpc and radius $R_{8.75}$=$R$/8$\arcmin$.75.

The present paper is an extension of the \textit{Chandra} study performed by GSH05 to the other regions of SNR G292.2--0.5 taking advantage of \textit{Chandra}'s excellent angular resolution ($\sim$0.5$\arcsec$ in the 0.5--10 keV energy range) and combining all the existing \textit{Chandra} data with the archived  \textit{XMM-Newton} observation of PSR~J1119$-$6127 taken in 2003 (Gonzalez et al. 2005). \textit{XMM-Newton}'s large collecting area ($\sim$4300 cm$^2$) makes it advantageous over \textit{Chandra} for studying the low-surface brightness emission from the SNR diffuse emission.  The main objectives of this study are to: 1) investigate the morphological differences between the SNR eastern and western sides, 2) determine the spectral and intrinsic properties of the SNR, and 3) estimate  the mass of the progenitor star that formed SNR G292.2$-$0.5. The latter goal is particularly motivated by the study of the connection between the high-magnetic field pulsars and magnetars. While on-going studies targeting this question focus on the compact objects themselves  (see e.g. Ng \& Kaspi 2011 and references therein), we here approach this question by examining the SNRs associated with these objects. In a follow-up paper, we address the progenitor mass of SNR~Kes~73 associated with a magnetar (a preliminary study was presented in Kumar et al. 2010).  

The paper is organized as follows: Sections 2 and 3 describe the observations and imaging analysis of the remnant, respectively. In Section 4, we present our spatially resolved spectroscopic study of the SNR using \textit{Chandra} and \textit{XMM-Newton}. In Section 5, we discuss the X-ray properties of the remnant including the evidence for ejecta, the interaction with a cloud, and presenting an estimate for the SNR intrinsic properties including its progenitor mass. Finally our conclusions are summarized in Section 6.

\section{Observations \& Data Analysis}
\label{2}

\subsection{\textit{Chandra}}
\label{2.1}

SNR G292.2--0.5 was observed with \textit{Chandra} on 2004 October 31--November 1 (ObsID: 4676) and 2004 November 2--3 (ObsID: 6153) for exposure times of 61.31 ks and 19.14 ks, respectively. The coordinates were positioned at the aimpoint of the back-illuminated S3 chip of the Advanced CCD Imaging Spectrometer (ACIS) onboard the \textit{Chandra X-ray Observatory}. These two observations of the eastern side of the SNR were combined with the previous \textit{Chandra} observation covering the western side and taken on 2002 March 31--April 1 (ObsID: 2833) for an exposure time of 57.55 ks. The results presented in this paper are based on the combined analysis of all the above mentioned observations. The CCD temperature was $-$120$^{\circ}$ C with a CCD frame readout time of 3.2 s in `TIMED' and `Very Faint' mode. The data were then reduced using the standard \textit{Chandra} Interactive Analysis of Observations (CIAO) version 4.3\footnote{http://cxc.harvard.edu/ciao.} routines. The level 1 raw event data were reprocessed to a new level 2 event file to remove pixel randomization and to correct for CCD charge transfer inefficiencies. The bad grades were filtered out and good time intervals were reserved. The resulting effective exposure times for the three observations after data processing are summarized in Table 1.

\begin{table}[h]
\caption{Summary of the effective exposure times for SNR G292.2--0.5}
\begin{tabular}{l l l l}
\hline\hline
Satellite &  ObsID & Detectors & Exposure \\
& & & time (ks) \\
\hline
\textit{Chandra} & 2833 & ACIS-S3 & 56.80 \\
\textit{Chandra} & 4676 & ACIS-S3 & 60.54 \\
\textit{Chandra} & 6153 & ACIS-S3 & 18.90\\
\textit{XMM-Newton} & 0150790101 & MOS1  & 48.33 \\
& & MOS2 & 49.77 \\
& & PN & 43.03 \\
 \hline
\end{tabular}
\end{table}

\begin{figure*}[ht]
\vspace{-0.5cm}
\hspace{4.75cm}\includegraphics[width=0.5\textwidth]{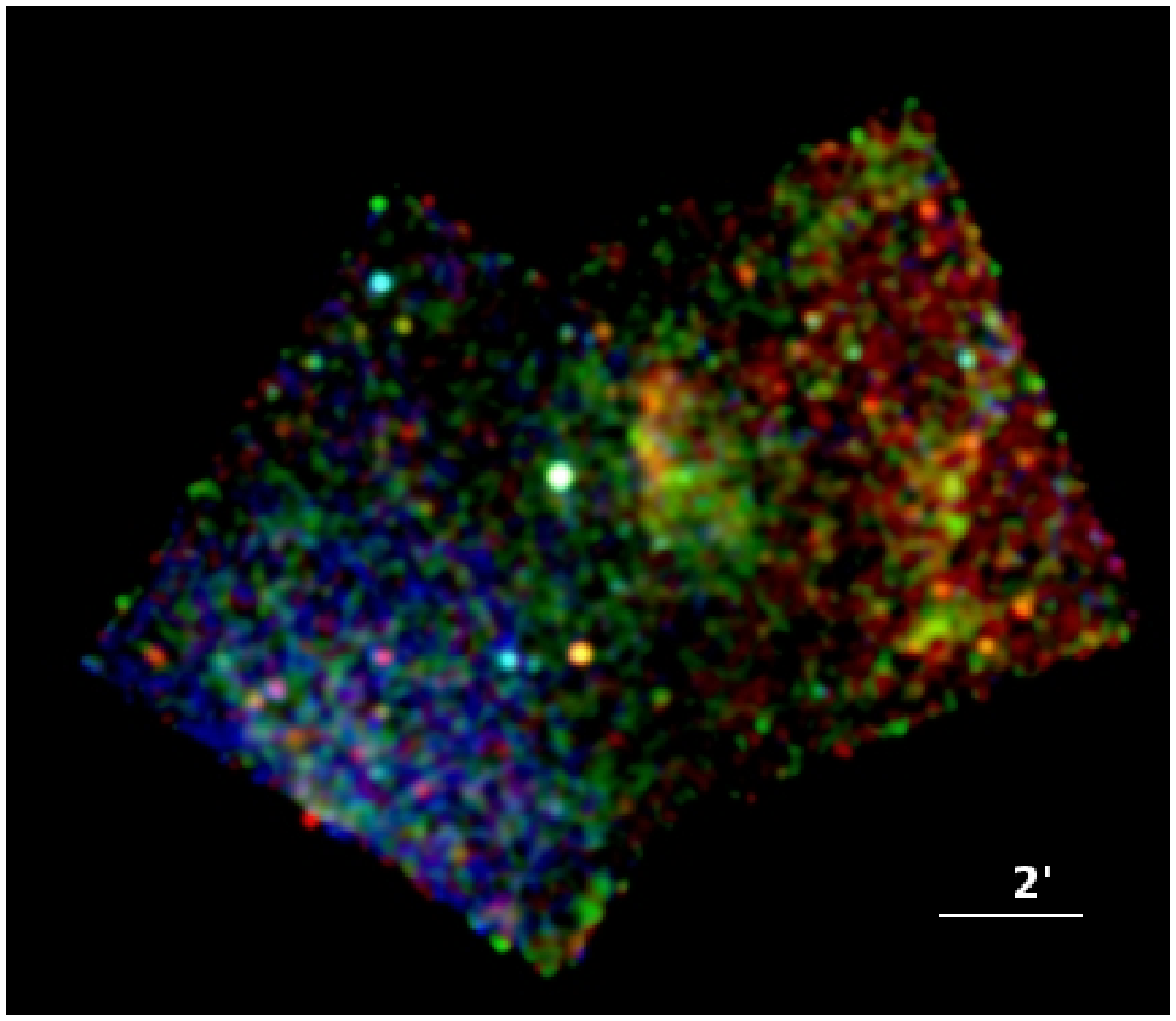}
\includegraphics[width=0.9\textwidth]{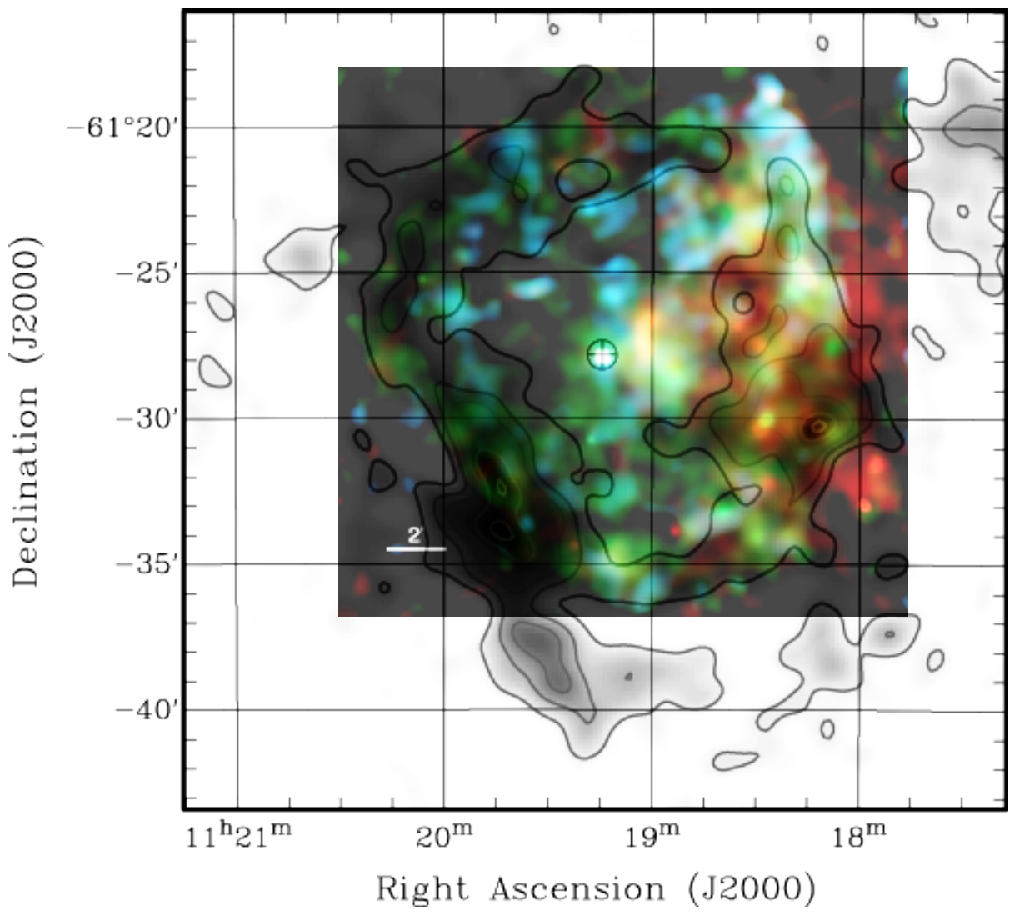}
\caption{(a) \textit{Top}: \textit{Chandra} ACIS-S3 tri-color exposure-corrected image of G292.2--0.5 combining observations of the eastern and western fields of the SNR.  (b) \textit{Bottom}: \textit{XMM-Newton} EPIC-MOS tri-color image overlaid with the radio contours from Caswell et al. (2004).  
The pulsar J1119$-$6127 is indicated by the white point-like source near the center.
For both images, red, green, and blue correspond to the 0.3--1.5, 1.5--3.0, and 3.0--10.0 keV energy range, respectively.}
\end{figure*}

\subsection{XMM-Newton}
\label{2.2}

The SNR G292.2--0.5 was observed with \textit{XMM-Newton} on 2003 June 26. The European Photon Imaging Camera (EPIC) PN (Struder et al. 2001) and MOS (Turner et al. 2001) cameras were operated using the medium filter. The PN camera was operated in the large window mode, allowing for a field-of-view of about 13$\arcmin$.5$\times$27$\arcmin$. In this mode, the southern side of the remnant was not covered by the PN chip. The MOS cameras were operated in the full frame mode and each covered the entire remnant. The data were analyzed with the \textit{XMM-Newton} Science Analysis System (SAS) software version 8.0.0\footnote{See http://xmm.esac.esa.int/sas/8.0.0} and the latest calibration files. Data from the contaminating background flares were excluded from our analysis.  The resulting effective exposure times for the MOS and PN detectors are shown in Table~1. 

\section{Imaging}
\label{3}

In order to investigate the morphological differences between the eastern and western sides of the SNR,  we performed an imaging analysis of the remnant in X-rays. 

In Figure~1 (a; top), we show the combined \textit{Chandra} ACIS-S3 tri-color image of the eastern and western sides of the SNR. The position of the pulsar is centered at $\alpha_{J2000}$=11$^{h}$19$^{m}$14$^{s}.26$ and $\delta_{J2000}$=$-$61$^{\circ}$27$\arcmin$49$\arcsec$.5 with 0.3$\arcsec$ error (Safi-Harb \& Kumar 2008). The data were divided into individual images in the soft (0.3--1.5 keV; \textit{red}), medium (1.5--3.0 keV; \textit{green}), and hard (3.0--10.0 keV; \textit{blue}) energy bands. The energy cuts were chosen so as to match the cuts for the \textit{XMM-Newton} image (see next). The resulting images were exposure-corrected and adaptively smoothed using a Gaussian with $\sigma$=1$\arcsec$--5$\arcsec$ for significance of detection 2 to 5. The individual exposure-corrected images were finally combined to produce the image shown in Figure~1 (a). 

Figure~1 (b; bottom) shows the combined MOS1+2 image of the remnant in the 0.3--1.5 keV (\textit{red}), 1.5--3.0 keV (\textit{green}), and 3.0--10.0 keV (\textit{blue}) bands. Following Gonzalez et al. (2005), individual MOS1/2 images were binned into pixels of  2$^{\prime\prime}$.5$\times$2$^{\prime\prime}$.5, which were then added and adaptively smoothed using a Gaussian with $\sigma$=5$\arcsec$--15$\arcsec$ to obtain a signal-to-noise ratio higher than 3$\sigma$.  The point sources (other than PSR~J1119$-$6127) were removed from the image. Background images (using the closed filter wheel data) and exposure maps at each energy band were created similarly and used to correct the final images. In order to compare the X-ray features with the remnant's radio morphology, we overlay on the MOS1+2 image the radio contours obtained from  Caswell et al. (2004). The resulting image is shown in Figure~1 (b). 

As mentioned earlier, the SNR G292.2-0.5 is mostly (although not fully, see Figure~1 (b)) covered by the large field of view of \textit{XMM-Newton}, unlike the \textit{Chandra} observations which focused only on either the eastern or the western side of the remnant. The compact PWN and a long southern jet are visible in the medium to hard X-ray band of the \textit{Chandra}  high-resolution image (see Figure 2 of Safi-Harb \& Kumar 2008 for a zoomed-on image of the PWN); \textit{XMM-Newton}'s 15$^{\prime\prime}$ half-power diameter  does not allow resolving the PWN.  For the SNR, the X-ray images reveal diffuse emission throughout most of the remnant, and the SNR does not appear limb-brightened in X-rays as in other shell-type SNRs. The western side is clearly characterized by a relatively bright and soft (red in color) diffuse emission, while the eastern side exhibits harder (blue in color) and fainter X-ray emission with a hint of shell-like emission visible towards the edge of the eastern field in the \textit{Chandra} image. We note here that the \textit{Chandra} image shows relatively more hard emission in the eastern side of the remnant in comparison to the \textit{XMM-Newton} image. This could be attributed to a combination of background+point source subtraction in the \textit{XMM-Newton} image and the fact that the 3.0--10 keV energy band of the \textit{XMM-Newton} observation suffered from a high degree of stray-light contamination on the instrument's mirrors from a nearby high-energy source. Therefore, as pointed out by Gonzalez et al. (2005), the detailed spatial distribution in this energy band should be examined with caution. A bright filamentary structure is also clearly visible in both the \textit{Chandra} and \textit{XMM-Newton} images inside the western side of the SNR, west of the pulsar (we refer to this region as `PSR--West' in the next section, see also Figure~2). The northern and southern regions of the remnant appear faint and featureless in X-rays. Bright diffuse emission is also seen towards the interior of the SNR along with several spatially resolved point sources as observed from the high-resolution \textit{Chandra} image (GSH05). 

The SNR G292.2--0.5 has been classified as a shell-type SNR by Crawford et al. (2001), with the radio image obtained by Caswell et al. (2004) clearly showing a bright shell-like structure surrounding the entire remnant and with an apparent thickness of $\sim$5$\arcmin$ in size (see Figure~1 (b)). Although, such a clearly defined shell is not evident in the X-ray images, the contours showing the radio shell overlap  with the bright X-ray shell on the western side. Furthermore, the radio images suggested a north-south elongated morphology for the remnant with the minor and major axes at 17$\arcmin$ and 18$\arcmin$, respectively (Crawford et al. 2001; Caswell et al. 2004).  Similarly, the \textit{XMM-Newton} image displays an elongated morphology, but with the major and minor axes at $\sim$20$\arcmin$ and 18$\arcmin$, respectively. The elongated bright feature extending in the northwest-southwest direction (seen more clearly in the \textit{XMM-Newton} image towards the western edge of the field of view) could be interpreted as the partial limb-brightened SNR shell. However, as evident in Figures~1 and 2, the current X-ray data shows more centrally bright and diffuse emission from within the remnant, suggesting that SNR G292.2--0.5 is rather a composite-type remnant. This classification is further supported by the X-ray detection of a PWN associated with PSR J1119$-$6127, which powers at least part of the X-ray emission seen in the SNR interior.

\section{Spatially resolved spectroscopy}
\label{4}

\begin{table*}[ht]
\center
\caption{Regions of diffuse emission extracted from the SNR G292.2--0.5 using \textit{Chandra} and \textit{XMM-Newton}}
\begin{tabular}{l l l l l}
\hline\hline
Region (Number)$^a$ & Right Ascension & Declination & Size & Net Counts$^b$\\
& h~m~s (J2000) & d m s (J2000) & & \\
\hline
SNR--East (1) & 11 19 35.087 & $-$61 31 34.04 & 3$\arcmin$.6$\times$1$\arcmin$.5 & 10660$\pm$377\\
SNR--West (4) & 11 18 26.666 & $-$61 28 01.41 & 1$\arcmin$.4$\times$2$\arcmin$.9 & 10720$\pm$275\\
PSR--West (3) & 11 18 59.184 & $-$61 28 04.96& 1$\arcmin$.3$\times$2$\arcmin$.2 & 10970$\pm$269\\
PSR--East (2)  &  11 19 23.652 & $-$61 29 09.00 & 1$\arcmin$.0$\times$3$\arcmin$.6 & 2828$\pm$260 \\
 \hline
\end{tabular}
\tablecomments{$^a$ The number in parenthesis besides the region name represents the region number as shown in Figure~2. \\
$^b$ Total \textit{Chandra} plus \textit{XMM-Newton} background-subtracted counts. For the \textit{Chandra} data we excluded counts above 7~keV due to poor signal-to-noise ratio. }
\end{table*}

The spectral analysis was performed using the X-ray spectral fitting package, XSPEC version 12.4.0\footnote{http://xspec.gsfc.nasa.gov.}, combining the \textit{Chandra} and \textit{XMM-Newton} data. For the \textit{XMM-Newton} data, the spectra were extracted in the 0.5--10 keV band, but for the \textit{Chandra} observations  we ignored the energy range above 7 keV due to poor signal-to-noise ratio. The contributions from point sources within the remnant were removed prior to the extraction of spectra. The background regions were selected from nearby source-free regions selected from the same CCD chip as the source (as shown in Figure~2). This has the advantage of removing any contamination from Galactic Ridge emission, and was possible since the emission from the SNR did not fill up the field of views of the \textit{Chandra} and \textit{XMM-Newton} observations.
The spectra were grouped by a minimum of 20 counts per bin\footnote{We note that we have experimented with different binning and our results are consistent with the fits presented here.}, and for each selected region they were fitted simultaneously in XSPEC. The errors quoted are at the 90$\%$ confidence level.  

For the spectral investigation, we defined four main elliptical regions from both the \textit{Chandra} and the \textit{XMM-Newton} observations. The regions along with their designated region numbers) are shown in Figure~2 and their properties are summarized in Table~2. The extracted regions are designated as follows: (a) SNR--East (Region 1): the outer diffuse emission seen towards the eastern side of the remnant and partially overlapping with the radio shell, (b) PSR--East (Region 2): the inner region east of the pulsar, (c) PSR--West (Region 3): the bright inner region seen west of the pulsar (this region was previously studied and designated as the ``eastern lobe'' by GSH05), and (d) SNR--West (Region 4): the outer diffuse emission seen towards the western side of the remnant and overlapping with the radio shell (as seen from Figure~1 (a)) and also studied by GSH05 where it was referred to as the ``western lobe". A few other SNR regions visible in the \textit{XMM-Newton} data, but not covered by the \textit{Chandra} observations,  were also examined. These regions, shown in Figure~2 (right), are named as follows: (i) SNR--Northwest (Region 5): the region to the far north-western side of the SNR, (ii) SNR--North (Region 6): the region in the northern interior of the remnant, and (iii) SNR--South (Region 7): the region to the south of the remnant. 

\begin{figure*}
\includegraphics[width=\textwidth]{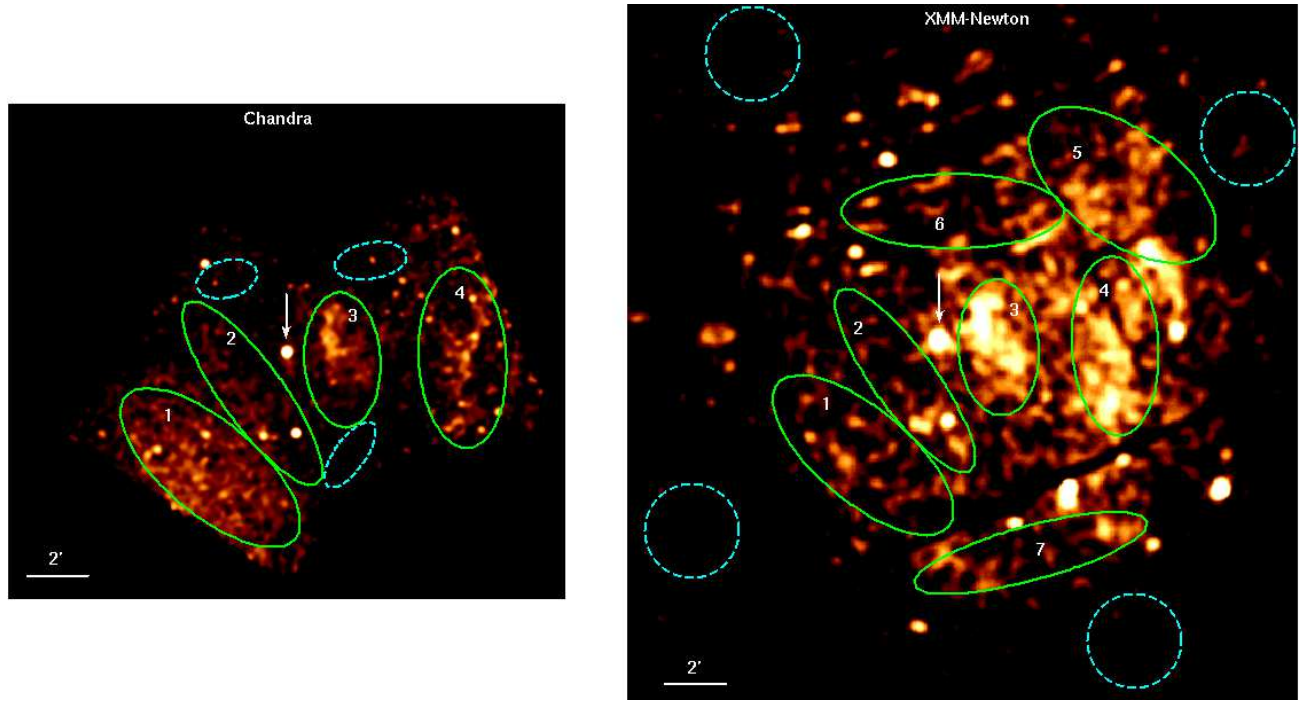}
\caption{(a) \textit{Left}: \textit{Chandra} intensity image of the SNR G292.2--0.5 (in logarithmic scale) 
smoothed using a Gaussian with $\sigma$=5$\arcsec$. Overlaid in green are the SNR diffuse emission regions selected for our spatially resolved spectroscopic study: SNR--East (region 1), PSR--East (region 2), PSR--West (region 3) and SNR--West (region 4). 
(b) \textit{Right}: \textit{XMM-Newton} MOS1 intensity image of SNR G292.2-0.5 (in logarithmic scale) smoothed using a Gaussian with $\sigma$=7$\arcsec$ showing the same regions selected for our spectroscopic study.  Three additional regions not covered with \textit{Chandra} were also selected: SNR--Northwest (region 5), SNR--North (region 6), and SNR--South (region 7). The position of the pulsar J1119$-$6127 is shown by a white arrow. Nearby background source-free regions selected from the same field of view, and closest to the source region, are
 shown as dotted circles/ellipses.}
\end{figure*}

The emission from the remnant was investigated using collisional ionization equilibrium (CIE) models such as MEKAL (Mewe et al. 1985;
Liedahl et al. 1995), Raymond-Smith (Raymond \& Smith 1977), and non-equilibrium ionization models (NEI) such as PSHOCK and VPSHOCK (Borkowski et al. 2001a) which are appropriate for describing the plasma in young SNRs. Furthermore, a power-law (PL) model  was also used to explore the high-energy continuum arising from any non-thermal emission from the SNR (as found by GSH05 in the western part of the remnant). 

Initially, the spectra from the extracted regions were fitted with the MEKAL model, which calculates the emission from an optically thin thermal plasma in CIE. The MEKAL fit resulted in a large reduced chi-squared $\chi^2_{\nu}$ $>$1.8 ($\nu$ being the number of degrees of freedom), which led us to rule out the CIE models -- a result that is also consistent with GSH05. Subsequently, we fitted the spectra using NEI models which resulted in better fits than the CIE models. 

NEI models such as PSHOCK and VPSHOCK (PSHOCK with varying abundances) are used to model plane-parallel shocks in a plasma, characterized by a constant electron temperature $(T)$ and an ionization timescale parameter $\tau$=$n_e t$, where $n_e$ is the post-shock electron density and $t$ is the time since the passage of the shock. While fitting the SNR regions using NEI models with variable abundances, the abundances of all elements were initially fixed at their solar values given by Anders \& Grevesse (1989), but varied as necessary to constrain the individual abundances. However, the abundances of H, He, C, and N were kept frozen to their solar values throughout, as their emission lines are below the lower limits of the detected energy range. Ni was tied to Fe throughout the spectral fitting and all the models included the Morrison \& McCammon (1983) interstellar absorption. Below, we describe the best fit spectral results for all the extracted diffuse emission regions.

\subsection{SNR--East (Region 1)}
\label{4.1}

We first fitted the SNR--East spectra with a one-component VPSHOCK model.  The fit yielded an interstellar absorption $N_H$=0.9$^{+0.1}_{-0.1}$$\times$10$^{22}$ cm$^{-2}$, $kT$=6.1$^{+2.2}_{-1.3}$ keV, $n_et$=1.1$^{+0.4}_{-0.3}$$\times$10$^{10}$ cm$^{-3}$~s,
and $\chi^2_{\nu}$=1.604 ($\nu$=951).  Here, we note that the temperature obtained from the fit is unusually high, even for a young remnant, indicating the presence of an additional component. A single PL component did not yield a better fit, as expected due to the presence of lines in the spectra. 

The fit was then further investigated by adding a second thermal component (owing to the expectation of a high- and low-temperature
plasma associated with the supernova blast wave and the reverse shock) or a second non-thermal PL component (assuming the possibility of an underlying synchrotron component arising from acceleration of particles at a shock). Though a two-component thermal model did not improve the fit or led to an unrealistically high-temperature in the case of a bremsstrahlung model, the addition of a non-thermal PL component to the VPSHOCK model improved the statistics ($\chi^2_{\nu}$=1.560, $\nu$=949) with an $F$-test probability of 
7.2$\times$10$^{-7}$. The best fit spectral parameters are summarized in Table~3. The value of $N_H$ obtained here is comparable to that obtained for PSR J1119--6127 ($\sim$1.8$^{+1.5}_{-0.6}$$\times$10$^{22}$ cm$^{-2}$ using a blackbody + PL model; Safi-Harb \& Kumar 2008).  Figure~3 (left) displays the best fit VPSHOCK+PL spectra of the SNR--East region.  We note that in this fit (and the fits for the other SNR regions described below), O was frozen to its solar value as allowing it to vary did not improve the fit significantly and led to a value consistent with solar or sub-solar values. The best-fit VPSHOCK+PL model required slightly enhanced abundances of Ne, Mg, and Si in SNR--East, hinting for the first time at the detection of shock-heated ejecta (further discussed in Section 5.1). 

\begin{figure*}[th]
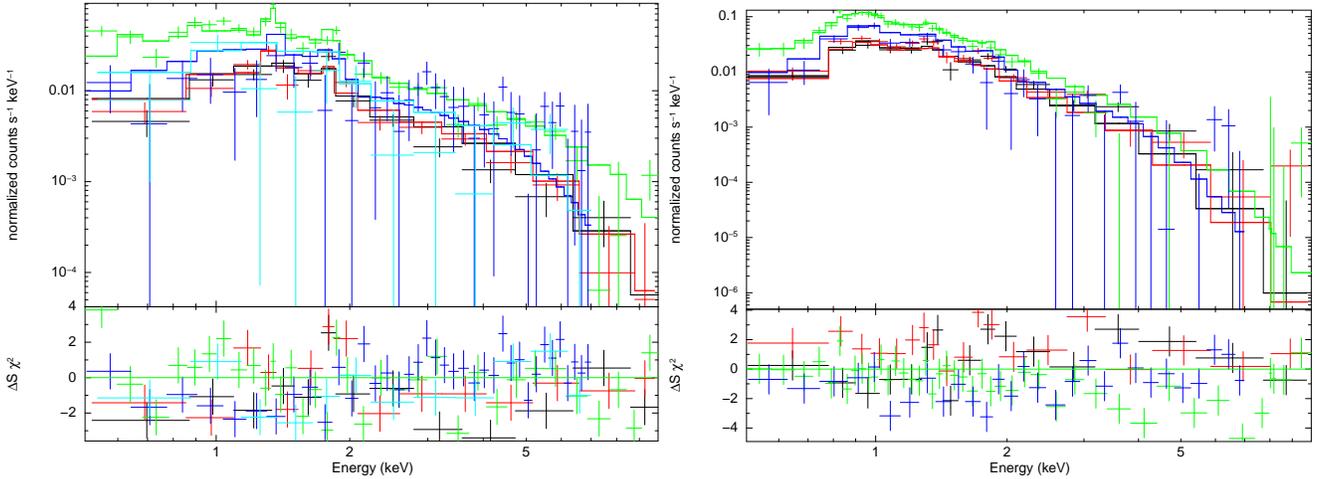

\includegraphics[width=0.35\textwidth,angle=-90]{f3a.eps}
\includegraphics[width=0.35\textwidth,angle=-90]{f3b.eps}
\caption{\textit{Chandra} and \textit{XMM-Newton} best-fit VPSHOCK+PL model for SNR--East (left) and SNR--West (right). The bottom panel displays the residuals in units of $\sigma$. PN: green, MOS1/2: black and red, \textit{Chandra}: blue and cyan. We note that the spectra have been rebinned for display purposes.}
\end{figure*}
\begin{figure*}
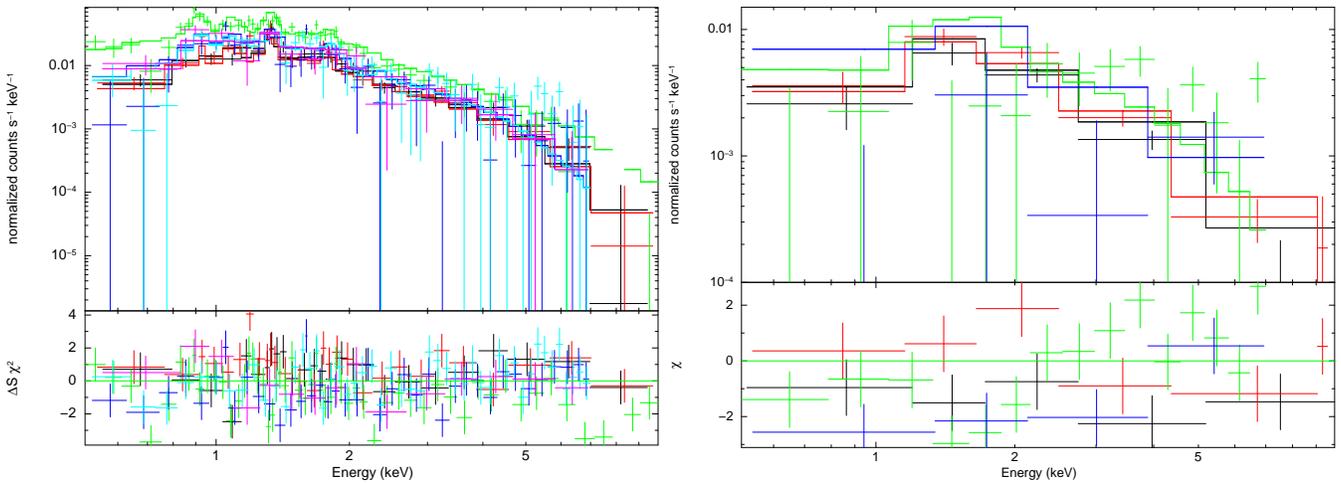

\includegraphics[width=0.35\textwidth,angle=-90]{f4a.eps}
\includegraphics[width=0.35\textwidth,angle=-90]{f4b.eps}
\caption{\textit{Chandra} and \textit{XMM-Newton} best fit VPSHOCK+PL model for PSR--West (left) and PSR--East (right). The bottom panel displays the residuals in units of $\sigma$. The colours are as in Figure 3, except for PSR--East which lacks PN data, green and blue refer to the \textit{Chandra} spectra. We note that the spectra have been rebinned for display purposes.}
\end{figure*}

\subsection{SNR--West (Region 4)}
\label{4.2}

The western side of the remnant, previously studied by GSH05 using \textit{Chandra} data taken in 2002, was best described by a PSHOCK model. Combining these data with the \textit{XMM-Newton} data, we find that the emission is described by a one-component VPSHOCK model (Figure 3, right) yielding the following values: $N_H$=1.0$^{+0.1}_{-0.6}$$\times$10$^{22}$ cm$^{-2}$,
$kT$=1.3$^{+0.3}_{-0.2}$ keV, $n_et$=3.6$^{+0.7}_{-0.6}$$\times$10$^{10}$ cm$^{-3}$ s, with $\chi^2_{\nu}$=1.433 ($\nu$=812). The elemental abundances were consistent with being solar or sub-solar (Table~3). Although the possibility of a second component was explored, the fits did not improve significantly. An $F$-test ruled out the necessity of a second thermal or non-thermal component. These results are consistent with, but better constrained than, those reported by GSH05 ($N_H$=0.5 (0.4--0.6)$\times$10$^{22}$ cm$^{-2}$, $kT$=1.6 (0.9--2.4) keV, and $n_et$=5.7 (3.3--12)$\times$10$^{10}$ cm$^{-3}$ s). 

\subsection{PSR--West (Region 3)}
\label{4.3}

The PSR--West spectra clearly revealed the presence of line emission, so were first examined using a VPSHOCK model. The spectral fit resulted in the following parameters: $N_H$=1.1$^{+0.1}_{-0.1}$$\times$10$^{22}$ cm$^{-2}$, $kT$=2.2$^{+0.2}_{-0.2}$ keV, $n_et$=4.4$^{+0.4}_{-0.4}$$\times$10$^{9}$ cm$^{-3}$ s, and $\chi^2_{\nu}$=1.161 ($\nu$=941). O and Si were frozen to their solar values (as allowing them to vary did not improve the fit significantly). For the other elemental abundances we obtained the following values: Ne=1.6$^{+0.3}_{-0.2}$, Mg=2.5$^{+0.4}_{-0.3}$, S=0.1$^{+0.7}_{-0.2}$, and Fe=1.8$^{+1.1}_{-0.8}$, indicating enhanced abundances and suggesting the detection of shock-heated ejecta. We also explored the possibility of a second thermal or non-thermal component. The addition of a PL component improved the fit with $\chi^2_{\nu}$=1.144 ($\nu$=939), yielding an $F$-test probability of 3.45$\times$10$^{-4}$.  The best fit spectral parameters for this region are summarized in Table 3 and the spectra are shown in Figure 4 (left). We note that the abundances for Mg and Si appear enhanced, and that the photon index is hard ($\Gamma$=1.5$^{+0.5}_{-0.2}$).
 
This region was also previously studied (referred to as the ``eastern lobe" by GSH05). A PSHOCK model fit to the \textit{Chandra} data alone gave an unusually high $kT$=9.1 (3.7--80)~keV, while a PL model fit gave a hard photon index $\Gamma$=1.5 (1.2--1.9). Our  combined \textit{Chandra} and \textit{XMM-Newton} fit, while it confirms the hard nature of the X-ray emission, allowed us to constrain the spectral properties and indicated the presence of thermal X-ray emission with slightly enhanced metal abundances, combined with non-thermal X-ray emission (as described above). The origin of this emission is further discussed in Section 5.4.

\subsection{PSR--East (Region 2)}
\label{4.4}

 To complement the study of PSR--West (section 4.3), we extracted an elliptical region from the eastern side of the pulsar (PSR--East) which appears fainter than the western side. We note that for this region we did not use PN since the X-ray emission fell on the chip edges. A VPSHOCK model fit resulted in the following values: $N_H$=0.9$^{+0.1}_{-0.1}$$\times$10$^{22}$ cm$^{-2}$, $kT$=5.5$^{+1.9}_{-1.1}$ keV, $n_et$=5.1$^{+0.8}_{-0.8}$$\times$10$^{9}$ cm$^{-3}$ s, and $\chi^2_{\nu}$=1.358 ($\nu$=510). The elemental abundances obtained are as follows: Ne=0.8$^{+0.3}_{-0.3}$, Mg=2.1$^{+0.9}_{-0.6}$, Si=1.2$^{+0.7}_{-0.5}$, and Fe=1.3$^{+1.4}_{-0.9}$. A single PL model fit yielded an $N_H$ value much lower than elsewhere in the remnant and showed residuals in both the soft and hard bands, ruling out this model and suggesting the need for an additional component.
 
The spectra were further explored using, like for PSR--West, a two-component VPSHOCK+PL model. The additional PL component was characterized by a hard photon index $\Gamma$=0.9$^{+0.6}_{-0.2}$, and the fit was slightly improved over the one-component VPSHOCK model, yielding $\chi^2_{\nu}$=1.345 ($\nu$=508, $F$-test probability of 0.027).  While the addition of the PL component did not improve the fit significantly, the fitted VPSHOCK model temperature has dropped to a more reasonable value of $kT$=1.9$^{+2.9}_{-0.8}$~keV which leads us to favor the two-component model. The best-fit parameters are summarized in Table~3.

\begin{table*}[ht]
\center
\caption{Best fit spectral parameters of SNR G292.2--0.5 using \textit{Chandra} and \textit{XMM-Newton.}}
\begin{tabular}{l l l l l}
\hline\hline Parameter & SNR--East (Reg. 1) & SNR--West (Reg. 4) & PSR--West  (Reg. 3) & PSR--East (Reg. 2) \\
\cline{2-5}
& VPSHOCK+PL & VPSHOCK  & VPSHOCK+PL & VPSHOCK+PL\\
\hline
 
  $ N_{H}$ ($10^{22}$ cm$^{-2}$) & 1.8$_{-0.4}^{+0.2}$ & 1.0$_{-0.6}^{+0.1}$ & 1.2$_{-0.2}^{+0.1}$ & 1.7$_{-0.4}^{+0.3}$\\

  kT (keV)& 2.3$_{-0.5}^{+2.9}$ & 1.3$_{-0.2}^{+0.3}$ & 2.1$_{-0.4}^{+0.4}$ & 1.9$_{-0.8}^{+2.9}$\\

 Ne  & 1.7$_{-0.6}^{+1.2}$ & 1 (frozen) & 1.9$_{-0.4}^{+0.5}$  & 1.3$_{-1.0}^{+1.3}$\\

 Mg  & 2.0$_{-0.4}^{+0.9}$ & 0.4$_{-0.1}^{+0.1}$ & 2.6$_{-0.6}^{+0.6}$ & 1.3$_{-0.6}^{+0.9}$\\

 Si  & 1.8$_{-0.5}^{+0.6}$ & 0.2$_{-0.1}^{+0.1}$ & 1 (frozen) & 1.4$_{-0.6}^{+0.7}$\\

 S  & 0.3$_{-0.3}^{+1.0}$ & 0.4$_{-0.4}^{+0.4}$ &  0.1$_{-0.1}^{+0.7}$ & 1 (frozen)\\

 Fe &  0.1$_{-0.1}^{+0.5}$ & 0.3$_{-0.1}^{+0.1}$ & 0.9$_{-0.6}^{+0.7}$ & 1.2$_{-0.6}^{+0.7}$ \\

 $n_et$ (cm$^{-3}$s) & 1.1$_{-0.2}^{+0.3}\times10^{10}$ & 3.6$_{-0.6}^{+0.7}\times10^{10}$ & 5.7$_{-0.7}^{+0.8}\times10^{9}$ & 7.5$_{-3.4}^{+2.5}\times10^{9}$\\

 $f_{unabs}$ (VP) & 3.6$_{-0.9}^{+2.4}\times10^{-12}$ & 7.0$_{-1.8}^{+1.6}\times10^{-12}$ & 5.3$_{-0.6}^{+1.0}\times10^{-12}$ & 3.2$_{-1.0}^{+0.4}\times10^{-12}$\\

$L_X$ (VP) & 3.0$^{+2.0}_{-0.8}$$\times$10$^{34}$ & 5.9$^{+1.4}_{-1.5}$$\times$10$^{34}$ & 4.5$^{+0.8}_{-0.5}$$\times$10$^{34}$ & 2.7$^{+0.3}_{-0.8}$$\times$10$^{34}$ \\
\hline

$\Gamma$ & 1.2$_{-0.1}^{+0.5}$ & \nodata & 1.5$_{-0.2}^{+0.5}$ & 0.9$_{-0.2}^{+0.6}$ \\

$f_{unabs}$ (PL) & 3.3$_{-0.7}^{+0.6}\times10^{-13}$ & \nodata & 0.9$_{-0.3}^{+0.2}\times10^{-13}$ & 3.0$_{-1.1}^{+0.7}\times10^{-13}$\\

$L_X$ (PL) & 2.8$^{+0.5}_{-0.6}$$\times$10$^{33}$ & \nodata & 0.8$^{+0.2}_{-0.3}$$\times$10$^{33}$ & 2.5$^{+0.6}_{-0.9}$$\times$10$^{33}$ \\
\hline

 $\chi_{\nu}^2 (dof)$ & 1.56 (949) & 1.43 (812) & 1.14 (939) & 1.34 (508) \\
\hline
\end{tabular}
\tablecomments{Errors are 2 $\sigma$ uncertainties.  All elemental abundances are in solar units given by Anders \& Grevesse (1989).
The 0.5--10 keV unabsorbed fluxes ($f_{unabs}$) and luminosities ($L_X$) quoted are in units of ergs cm$^{-2}$ s$^{-1}$ and ergs s$^{-1}$, respectively.
VP is VPSHOCK, PL is power-law.}
\end{table*}

\subsection{Other regions of the remnant}
\label{4.5}

An attempt has been made to study a few other SNR regions visible only in the \textit{XMM-Newton} data (see Figure~2): SNR--Northwest (region 5), SNR--North (region 6) and SNR--South (region 7). However, as further described below, the relatively small number of counts extracted from SNR--North and SNR--South did not allow us to constrain the parameters as for the other regions.

 The northwest region of the remnant is extracted from an elliptical region of size 1$\arcmin$.8$\times$3$\arcmin$.5 ($\alpha_{J2000}$=11$^h$18$^m$24$^s$.825 and $\delta_{J2000}$=$-$61$^{\circ}$22$\arcmin$50$\arcsec$.01), yielding 7128$\pm$156 background-subtracted counts in the 0.5--10 keV range. The spectra of the SNR--Northwest region were first fitted by a one-component VPSHOCK model which yielded the following parameters:  $N_H$=1.0$^{+0.2}_{-0.2}$$\times$10$^{22}$ cm$^{-2}$, $kT$=2.5$^{+1.4}_{-0.4}$ keV, $n_et$=5.9$^{+0.8}_{-0.7}$$\times$10$^{9}$ cm$^{-3}$ s, and $\chi^2_{\nu}$=1.291 ($\nu$=653). The metal abundances obtained are consistent with solar values: Ne=0.9$^{+0.3}_{-0.2}$, Mg=1.3$^{+0.4}_{-0.3}$, and Si=0.5$^{+0.3}_{-0.3}$. Adding a PL component reduced the fitted VPSHOCK temperature and improved the fit, yielding $\chi^2_{\nu}$=1.223 (651) ($F$-test probability=7.6$\times$10$^{-9}$) and  the following parameters: $N_H$=1.3$^{+0.1}_{-0.2}$$\times$10$^{22}$~cm$^{-2}$, $kT$=0.9$^{+0.3}_{-0.2}$ keV, $n_et$=0.6$^{+0.2}_{-0.2}$$\times$10$^{10}$ cm$^{-3}$~s, and $\Gamma$=0.9$^{+0.2}_{-0.3}$. The metal abundances were consistent with solar values and the non-thermal component contributed only $\sim$3\% of the total unabsorbed flux.

SNR--North (region 6) was extracted from an elliptical region of size 3$\arcmin$.6$\times$1$\arcmin$.2 centered at
$\alpha_{J2000}$=11$^h$19$^m$10$^s$.553 and $\delta_{J2000}$=-61$^{\circ}$23$\arcmin$41$\arcsec$.57, yielding 3216$\pm$146 
background-subtracted counts in the 0.5--10 keV range.  The emission from this region was best
fitted with a PSHOCK model yielding an $N_H$=0.8$^{+0.2}_{-0.2}$$\times$10$^{22}$ cm$^{-2}$, $kT$=8.9$^{+4.9}_{-2.6}$ keV,
$n_et$=2.0$^{+1.1}_{-0.7}$$\times$10$^{9}$ cm$^{-3}$ s, global abundance=0.9$^{+4.9}_{-0.9}$, and $\chi^2_{\nu}$=1.241 ($\nu$=178). We note that the temperature derived from this fit is very high and hence, we further explored the fit using a second PL component. However, the parameters could not be well constrained. A single PL model gave a very low $N_H$ and was also ruled out based on the residuals. The poor statistics did not allow us to explore the fits further.

SNR--South (region 7) was extracted from an elliptical region of size 3$\arcmin$.9$\times$0$\arcmin$.9, centered at $\alpha_{J2000}$=11$^h$18$^m$49$^s$.505 and $\delta_{J2000}$=$-$61$^{\circ}$34$\arcmin$43$\arcsec$.90, yielding only 1303$\pm$86 background-subtracted counts in the 0.5--10 keV range. We note here that this region was not covered by PN and so we only used the MOS1/2 data. As in the other regions, we examined the fit using a PSHOCK and a PSHOCK+PL model. The  PSHOCK fit provided the following values: $N_H$=0.6$^{+0.2}_{-0.2}$$\times$10$^{22}$ cm$^{-2}$, $kT$=3.0$^{+1.6}_{-0.9}$ keV, $n_et$=9.9 ($>$ 5.5)$\times$10$^{8}$ cm$^{-3}$ s, global abundance=1.0$^{+0.2}_{-0.1}$, and $\chi^2_{\nu}$=1.036 ($\nu$=155).  The two-component PSHOCK+PL model provided an acceptable fit with the following values: $N_H$=1.4$^{+0.8}_{-0.1}$$\times$10$^{22}$ cm$^{-2}$, $kT$=1.9$^{+1.8}_{-1.1}$ keV, $n_et$=9.2$^{+19.8}_{-5.5}$$\times$10$^{10}$ cm$^{-3}$~s, global abundance=0.1$^{+1.0}_{-0.5}$,  $\Gamma$=1.4$^{+0.5}_{-0.3}$, and $\chi^2_{\nu}$=1.060 ($\nu$=153). Given the relatively low number of counts collected from this region, the errors were here computed by freezing the other parameters. Statistically, a second component for this region was not required, but comparing the fit parameters with those obtained for the other regions of the remnant (for example, the ionization timescale and the column density for SNR--South using a one-component model are much smaller than those obtained for all the other regions, and the temperature obtained was higher with a single component model), we favour the two-component model fit for SNR--south. However, the one-component model fit could also imply expansion of the SNR into a more tenuous medium in the south leading to hotter plasma, lower ionization timescale, and fainter X-ray emission. Additional observations are needed to confirm the spectral nature of the emission from these regions.

\section{Discussion}
\label{5}

We have performed a spatially resolved spectroscopic study of the diffuse emission regions within SNR G292.2--0.5 using
\textit{Chandra} and \textit{XMM-Newton}. The emission from most regions (shown in Figure~2) can be described by a two-component thermal+non-thermal (VPSHOCK+PL) model, except for SNR--West which required only a one-component thermal (VPSHOCK) model.  In all regions, the fitted temperature is high ($kT$$\geq$1~keV) and the ionization timescale is low ($n_et$$<$10$^{11}$~cm$^{-3}$~s), indicating that the plasma is hot and has not yet reached ionization equilibrium.  As well the PL photon index is hard ($\Gamma$$\sim$1--2), much flatter than that observed in non-thermal SNRs and consistent with the photon index observed in the compact PWN associated with 
PSR J1119--6127.  Below, we discuss these results based on our best-fit spectral parameters summarized in Table~3.
We stress here that while the discussion of the SNR properties is based on these tabulated parameters, we have experimented with a number of  available XSPEC models appropriate for modelling plasma in young SNRs (such as $vnei$) and they provide similar global parameters. 
The model fits presented here best reproduce the main features observed in our X-ray spectra. More complex and multi-component models are needed to provide a more robust description of this complex SNR.  Such a study will have to await improved models as well as better quality data.

\subsection{Evidence of ejecta}
\label{5.1}

The X-ray emission from young SNRs is characterized by the propagation of the forward shock/blast wave into the ISM and the reverse shock running back into the ejecta. Both shocks can be currently resolved and studied with missions like \textit{Chandra} and \textit{XMM-Newton}. We next discuss the spectral properties and morphology of the remnant in the light of the possible identification of ejecta for the first time.

In Figure~1 (b), we show the contours of the radio shell overlaid on the X-ray image obtained with \textit{XMM-Newton}. The outer radio shell presumably identifies the supernova blast wave that is expanding into and shocking the surrounding medium. 
Comparing with Figure~2, the X-ray emission from regions SNR--West and SNR-Northwest falls within the boundaries of the radio shell. Both of these regions are fitted by a VPSHOCK model with comparable temperatures ($kT$$\sim$1~keV), ionization timescales ($n_et$$\sim$10$^{10}$~cm$^{-3}$~s), and solar (or sub-solar) abundances. This suggests that the X-ray emission in those regions arises mainly from shocked interstellar or circumstellar medium. In comparison, the regions SNR--East and PSR--West\footnote{The PSR--East spectrum, while it reveals very slight enhancement of metal abundances over the solar values, suffered from poor statistics therefore prohibiting us from constraining the abundances.}, 
reveal enhanced metal abundances for Ne, Mg and Si, hinting at the first evidence of reverse-shocked ejecta associated with SNR G292.2--0.5.  Moreover,  the inferred ionization timescales from regions SNR--East, PSR--East, and PSR--West range from 5.7$^{+0.8}_{-0.7}$$\times$10$^9$ cm$^{-3}$~s for PSR--West to 1.1$^{+0.3}_{-0.2}$$\times$10$^{10}$ cm$^{-3}$~s for SNR--East, lower than that in the SNR--West region (3.6$^{+0.7}_{-0.6}$$\times$10$^{10}$~cm$^{-3}$~s). This further indicates that the plasma in the outermost region, SNR--West, was shocked earlier than the remnant's more interior regions.  As well, the inferred temperature of the more interior regions is $\sim$2~keV, hotter than in SNR--West. These results further support the conclusion that the X-ray emission from SNR--West likely originates from the blast wave, while the more interior regions correspond to (at least partly) reverse shock-heated ejecta expanding in a lower density medium.

\subsection{X-ray properties of the SNR G292.2--0.5}
\label{5.2}

In the following, we derive the X-ray properties of SNR G292.2--0.5 shown in Table~4 using the VPSHOCK model parameters summarized in Table 3. In our calculations, we take the radius of the SNR as $\sim$8$\arcmin$.75 (Caswell et al. 2004), which translates to a physical size of $R_s$=21.4$D_{8.4}$ pc=6.6$\times$10$^{19}$ $D_{8.4}$ cm. 

The volume of the X-ray emitting region ($V$) is estimated by assuming that the plasma fills an ellipsoid with the length and width equivalent to those of the extracted SNR regions (Table~2) and the depth taken as the radius of the remnant. The emission measure $EM$=$\int n_en_HdV$$\sim$$fn_en_HV$ is defined as a measure of the amount of plasma available to produce the observed flux,  where $n_e$ is the post-shock electron density for a fully ionized plasma, $n_H$ is the mean Hydrogen density, and $f$ is the volume filling factor of the hot gas. For cosmic abundance plasma and the strong shock Rankine-Hugoniot jump conditions, the ambient density $n_0$ can be estimated from the electron density $n_e$ as:  $n_e$=4.8$n_0$;  $n_0$ here includes only Hydrogen (see e.g. Safi-Harb et al. 2000). The emission measure values are estimated from the normalization  $K$ obtained from the X-ray spectral fits as $K$=($10^{-14}$/4$\pi$$D^2$)$\int$$n_en_HdV$. Using the above equations, we subsequently calculated the post-shock electron density ($n_e$) and the ambient density ($n_0$) of the X-ray  emitting plasma, as summarized in Table 4. The low inferred ambient density of $n_0$$\sim$(1--2)$\times$10$^{-2}$~$f^{-1/2}D^{-1/2}_{8.4}$~cm$^{-3}$ suggests that the SNR is mostly expanding into a low-density medium. This is further discussed in Section 5.5.

We now discuss the age of the SNR.  A lower estimate can be inferred assuming the SNR is still in its free expansion phase. For an initial expansion velocity of $v_0$$\sim$5000 km~s$^{-1}$ (which is reasonable for a core-collapse explosion; see Reynolds 2008), the SNR's size implies an age of
$\gtrsim$4.2~$D_{8.4}$~kyr, a factor of $\sim$2.2 higher than the pulsar's age upper limit of 1.9~kyr (see Section 1).
We note that this derived age is appropriate under the assumption of an explosion into a wind bubble blown by the progenitor with the SNR shock only fairly recently reaching the edge of the bubble and being slowed to its current velocity.
Requiring the SNR's age to be $\leq$1.9~kyr (current pulsar's age upper limit) necessitates an expansion velocity of $\geq$11200~$D_{8.4}$~km~s$^{-1}$, which is unusually high (at the assumed distance of 8.4~kpc) for the expansion of a blast wave of a core-collapse explosion.

Before we further address the apparent discrepancy between the SNR and the pulsar ages, we now assume a Sedov phase of evolution for the SNR. While the SNR may not have yet fully entered this phase, this assumption gives  an upper limit on the SNR age.
An SNR enters the Sedov phase when the swept-up mass becomes comparable to the ejected mass, and settles fully in that phase once the swept-up mass exceeds the mass of the ejecta by a factor of $\geq$10. Under the assumption of a uniform ambient medium, the swept-up mass is estimated as\footnote{While this mass is comparable to the progenitor star's mass estimated in Section 5.5, the ejected mass is believed to be considerably smaller as the star should have lost a significant fraction of its mass before exploding (see details in Section 5.5).}  $M_{sw}$=1.4$m_pn_0$$\times$($\frac{4}{3}$$\pi$$R_s^3f$)=33$^{+7}_{-10}$~$f^{1/2}D^{5/2}_{8.4}$~$M_{\sun}$, consistent with a massive progenitor and an evolutionary stage between ejecta-dominated and Sedov.

The age of the remnant is given by $t_{SNR}$=$\eta$$R_s$/$V_s$ where $\eta$=0.4 for a blast wave expansion in the Sedov phase (Sedov 1959). The shock velocity can be estimated as $V_s$=(16$k_BT_s$/3$\mu$$m_H$)$^{1/2}$ assuming full equilibration between the electrons and ions (Sedov 1959), where $\mu$=0.604 is the mean mass per free particle for a fully ionized plasma, $k_B$=1.38$\times$10$^{-16}$ ergs~K$^{-1}$ is Boltzmann's constant, and the post-shock temperature $T_s$ is related to the temperature inferred from the X-ray fit ($T_X$) as $T_X$$\sim$1.27$T_s$ in the Sedov phase (since the temperature increases inward behind the shock radius; Rappaport et al. 1974).  Using the spectral fit values obtained for SNR--West and assuming that its emission characterizes the blast-wave component (see Section 5.1), we estimate a shock velocity of $V_s$=(1.1$^{+0.2}_{-0.1}$)$\times$10$^3$~km~s$^{-1}$ for the forward shock, and therefore, an age of $t_{SNR}$ (Sedov)=7.1$^{+0.5}_{-0.2}$ $D_{8.4}$ kyr for SNR G292.2--0.5. We note here that the shock ages inferred from ionization timescales ($n_et$) indicate a range of shock ages (time since the passage of shock) of $t$=(1.8$^{+0.3}_{-0.3}$--10$^{+3}_{-3}$)$f^{1/2}$$D^{1/2}_{8.4}$~kyr. These values (Table~4) are consistent with the remnant's age, noting that $f$ can have different values across the remnant. For approximately similar values of $f$ for the four SNR regions shown in Table~4, the lower shock age for the more interior regions supports a reverse-shock origin since the reverse shock develops later in the SNR evolutionary stage.

Assuming Sedov dynamics, the explosion energy of the remnant can be estimated as $E_0$=(1/2.02)$R_s^5m_nn_0t_{SNR}^{-2}$=0.6$^{+0.1}_{-0.2}$$\times$10$^{51}$ $f^{-1/2}$$D^{5/2}_{8.4}$ ergs, where $m_n$=1.4$m_p$ is the mean mass of the nuclei. We note here that the explosion energy will be an underestimate and the remnant's age will be an overestimate if full electron-ion equilibration in the shock has not yet been achieved in G292.2--0.5, since the electron temperature (and hence, $V_s$) is a lower limit to the shock temperature. We note however that a $VNPSHOCK$ model\footnote{The VNPSHOCK model in XSPEC is a plane-parallel shock plasma model with separate ion and electron temperatures.} fit to the X-ray data yielded equal electron and ion temperatures. The total X-ray luminosity of the four SNR extracted regions is $L_X$ (0.5--10 keV)=4$\pi$$D^2_{8.4}$$f_{unabs}$=1.7$\times$10$^{35}$~$D^2_{8.4}$~ergs~s$^{-1}$. This should be considered as a lower limit to the SNR X-ray emission (see Section 4.5).

In summary,  the SNR age is estimated to be between 4.2~$D_{8.4}$ kyr (free expansion phase, assuming an expansion velocity of 5000 km~s$^{-1}$) and  7.1~$D_{8.4}$~kyr (Sedov phase), a factor of a few higher than the pulsar's age upper limit. Discrepancies between between pulsar and SNR ages have been observed in other SNRs (see e.g. Arzoumanian et al. 2011). The pulsar's age, $\tau$, is highly dependent on its initial spin period ($P_0$) and braking index ($n$), and can be expressed as follows: $\tau$=$\frac{P}{(n-1)\dot{P}}$  [1-$\left(\frac{P_0}{P}\right)^{(n-1)}]$;
where $P$ is the current spin period. The pulsar's characteristic age is $\tau_c$=$\frac{P}{2\dot{P}}$, which corresponds to $n$=3 and $P_0$$<<$$P$.
As mentioned earlier, the braking index  for PSR~J1119$-$6129 has been measured to be $n$=2.684$\pm$0.002 using more than 12 yr of timing data (Weltevrede et al. 2011), implying an age of 1.9~kyr under the assumption of $P_0$$<<$$P$ (=408 ms). In Figure~5, we plot the pulsar's true age as a function of the (unknown) initial period, $P_0$ for various values of $n$ (1.5, 2.0, 2.7, 3.0, 3.5, and 4.0). Corresponding upper limits on the pulsar's true age are: 6.6, 3.3, 1.9 (current upper limit), 1.6 (characteristic age), 1.3 and 1.1 kyr, respectively.
Therefore to reconcile the pulsar's age with the SNR's age estimate, the braking index has to be $<$ 2.0 for most of the pulsar's lifetime
and must have increased to $\sim$2.7 only recently; as well its initial spin period should be much smaller than its current period ($P_0$$<$$<$$P$). 
We note here that the recent radio timing analysis of the pulsar has shown some unusual Rotating Radio Transient (RRAT)-like behaviour (Weltevrede et al. 2011). Such erratic RRAT-like events and the unusual timing characteristics observed for PSR~J1119$-$6127  may also partially account for the uncertainties in the pulsar's age measurement. 
 
\begin{figure*}
\center
\includegraphics[width=0.8\textwidth]{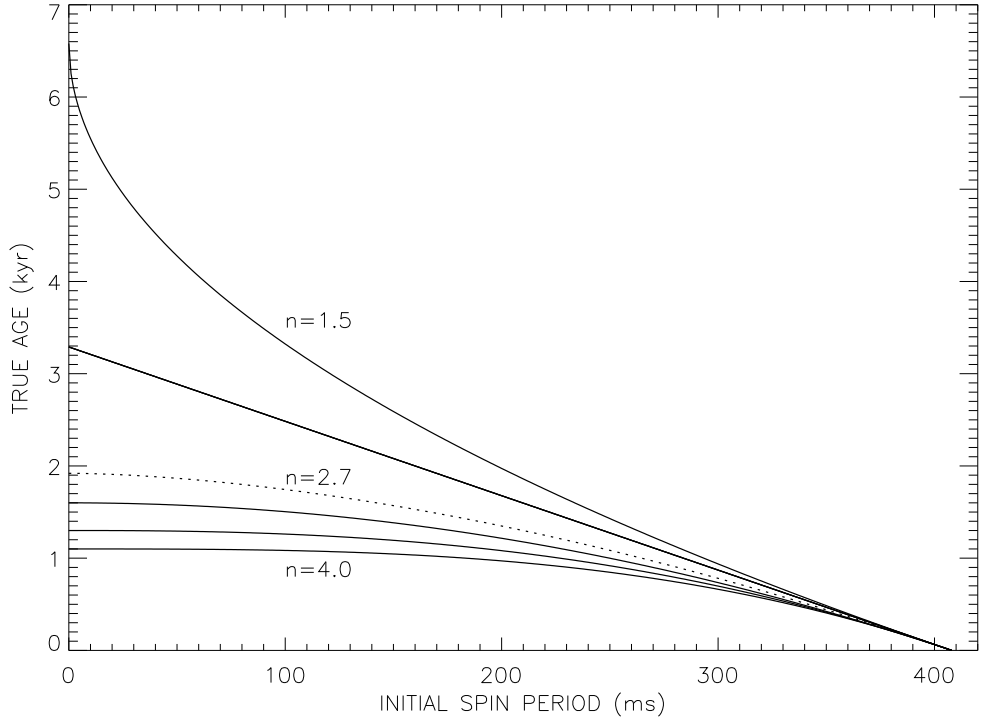}
\caption{The true age of PSR J1119$-$6127 as a function of its initial spin period, for different values of the braking index ($n$): top to bottom curves correspond to $n$=1.5, 2.0, 2.7 (dotted line, recently measured braking index), 3.0, 3.5 and 4.0. Corresponding upper limits on the pulsar age are: 6.6, 3.3, 1.9, 1.6, 1.3 and 1.1 kyr, respectively. The current pulsar's period is 408 ms.  See Section 5.2 for details.}
\end{figure*}

\begin{table*}
\center
\caption{Derived X-ray parameters of SNR G292.2--0.5}
\begin{tabular}{l l l l l}
\hline\hline
Parameters & SNR--East & PSR--East & PSR--West & SNR--West\\
\hline

Emission measure, EM (10$^{56}$~$fD^{-2}_{8.4}$~cm$^{-3}$) & 2.7$^{+1.0}_{-0.6}$ & 2.4$^{+0.3}_{-0.4}$ & 3.5$^{+0.4}_{-0.3}$ & 6.7$^{+1.5}_{-1.7}$ \\

Electron density, $n_e$ (10$^{-1}$~$f^{-1/2}$$D^{-1/2}_{8.4}$~cm$^{-3}$) & 0.6$^{+0.2}_{-0.1}$ & 0.7$\pm$0.1 & 1.0$\pm$0.1 & 1.1$^{+0.2}_{-0.3}$ \\

Ambient density, $n_0$ (10$^{-2}$~$f^{-1/2}$$D^{-1/2}_{8.4}$~cm$^{-3}$) & 1.0$^{+0.4}_{-0.2}$ & 1.0$\pm$0.2 & 2.0$\pm$0.2 & 2.0$^{+0.4}_{-0.6}$ \\

Shock age, $t$ ($f^{1/2}$$D^{1/2}_{8.4}$ kyr) & 5.6$^{+2.4}_{-1.4}$  & 3.4$^{+1.6}_{-1.2}$ & 1.8$^{+0.3}_{-0.3}$ & 10$\pm$3 \\
 \hline
\end{tabular}
\end{table*}

\subsection{Association with a molecular cloud?}
\label{5.3}

As discussed earlier under Section 3, the eastern and western sides of SNR G292.2--0.5 appear different in morphology. It has been suggested that asymmetry and differences within an SNR can be observed if the ISM is not isotropic (Cox et al. 1999) or if the SNR is interacting with a molecular cloud (Chevalier 1999). P01 had already suggested the presence of a dark cloud DC 292.3--0.4, catalogued by Hartley et al. (1986) from a visual inspection of the ESO/SERC Southern $J$ Survey plates, towards the eastern side of the SNR with a diameter of $\sim$16$\arcmin$ and an estimated contribution to $N_H$ of $\sim$(6--7)$\times$10$^{21}$ cm$^{-2}$. This is further supported by our imaging analysis (Figure~1), which shows a lack of soft X-ray emission from the eastern side of the SNR, suggesting the presence of additional absorbing material or a cloud absorbing the soft X-rays in that region. Furthermore, our spatially resolved spectroscopic study clearly indicates a higher column density in the eastern side (see Table~3), with $N_H$=1.8$^{+0.2}_{-0.4}$$\times$10$^{22}$ cm$^{-2}$ for SNR--East versus 1.0$^{+0.1}_{-0.6}$$\times$10$^{22}$ cm$^{-2}$ for SNR--West.  We estimate that this absorbing material should be contributing an $N_H$ of $\sim$8$\times$10$^{21}$ cm$^{-2}$, the difference in $N_H$ between SNR--East and SNR--West, which is consistent with that inferred for the dark cloud (P01).  

We further investigate the above scenario by inspecting the region surrounding SNR G292.2--0.5. Figure~6 (a) shows the optical intensity image obtained from ESO Digitized Sky Survey (DSS\footnote{http://archive.eso.org/dss/dss}) and smoothed using a Gaussian of $\sigma$=5$\arcsec$. Overlaid on the image (green circle) is the catalogued dark cloud DC 292.3--0.4 (Harley et al. 1986)\footnote{No further information (e.g. on the morphology) besides the location and approximate extent was given on this cloud.}, centered at $\alpha_{J2000}$=11$^{h}$20$^{m}$25$^{s}.0$ and $\delta_{J2000}$=$-$61$^{\circ}$22$\arcmin$00$\arcsec$ and with a radius of 8$\arcmin$. The catalogued size clearly exceeds the extent of the cloud seen in the DSS image (see also Figure 6 (b)).  In the same image, we also identify another dark region (shown by a green ellipse towards the centre of the image) centered at approximately $\alpha_{J2000}$=11$^{h}$19$^{m}$19$^{s}.1$ and $\delta_{J2000}$=$-$61$^{\circ}$31$\arcmin$03$\arcsec$.0 with a size of $\approx$6$\arcmin$.5$\times$2$\arcmin$.2, and which appears to overlap with SNR--East and most of PSR--East (regions 1 and 2, respectively, shown as white ellipses). The small purple cross in Figure~6 (a) marks the position of PSR J1119-6127. In Figure~6 (b), we overlay, in red, the corresponding dark cloud contours from the DSS image on the \textit{XMM-Newton} MOS (black contours) and \textit{Chandra} (blue contours) X-ray images. We speculate that the dark elliptical region in the south-eastern field of G292.2$-$0.5 likely represents another uncatalogued dark cloud; alternatively it could be a previously unidentified substructure of DC 292.3--0.4 since the cloud size reported by Hartley et al. (1986) does not accurately reflect the shape of the cloud seen in the DSS image. This is corroborated by the detection of two CO features with line of velocities of $-$12.6~km~s$^{-1}$ and $-$1.6~km~s$^{-1}$ towards the centre of DC 292.3--0.4 (Otrupcek et al. 2000). While its not clear whether both features are associated with the same cloud to the north-east (catalogued by Hartley et al. 1986), it is possible that one of them could be associated with the cloud we have identified coinciding with the south-eastern side of the remnant. To the best of our knowledge,  there is no CO observation targeted to study this dark cloud.  As well, we are not aware of any recent CO study of the environment of G292.2$-$0.5. Nevertheless, the presence of these dark clouds towards the eastern part of the remnant and the detected CO features support the presence of additional foreground absorbing material that would explain the lack of soft X-ray emission from that region. Furthermore, the dark cloud largely obscuring PSR--East  (Figure~6) could further explain the lack of bright emission from PSR--East in comparison with the emission from PSR--West (Figure~2).  

Next, we discuss the observed morphology and the non-thermal emission from the eastern side of the remnant in the light of a possible interaction with a nearby cloud. The presence of an adjacent cloud can affect the evolution and appearance of a SNR as suggested e.g. by Wardle \& Yusef-Zadeh (2002). The expanding SNR is slowed at the location of the cloud, causing it to appear distorted. The reflected shock from the cloud moves into the interior of the remnant and interacts with the SNR-driven shock waves. This results in heating the cloud material and filling the interior of the SNR with gas, which can substantially modify the X-ray emission and lead to a composite-type morphology  (Rho \& Petre 1999). The hot thermal X-ray emission detected from the SNR interior supports this scenario. Such an interaction however is expected to lead to OH maser emission (Wardle \& Yusef-Zadeh 2002), but to the best of our knowledge no maser emission has been detected from G292.2--0.5. 

Furthermore, acceleration of particles to very high energies at the shock-cloud interface can cause synchrotron
radiation observed at X-ray wavelengths, as e.g. in the case of SNR RCW~86 (Borkowski et al. 2001b). The spectrum of SNR--East is indeed best fitted by a thermal+non-thermal model;  the non-thermal component could then be interpreted as arising from the interaction of the SN shock wave with a nearby molecular cloud. Morphologically, this is supported by the correspondence between the hard X-ray emission from SNR-East and the radio shell (see Figure~1). We note however that the \textit{XMM-Newton} and \textit{Chandra} observations do not fully cover the radio shell. As well,  the non-thermal X-ray flux is only  $\sim$10\% of the thermal flux, and no TeV emission has been detected from that region (Djannati et al. 2009, see next section) as observed in other interacting SNRs. Furthermore, the hard power-law photon index ($\Gamma$=1.2$^{+0.5}_{-0.1}$) inferred for SNR--East is unusually hard and puzzling. More sensitive X-ray observations with coverage of the radio shell, as well as CO observations towards the eastern side of the remnant, are needed to confirm or rule out any interaction with a molecular cloud.

\begin{figure*}[ht]
\vspace{-0.5cm}
\hspace{4cm}\includegraphics[width=0.7\textwidth]{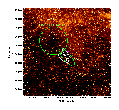}
\center
\vspace{-1.9cm}
\includegraphics[width=0.7\textwidth]{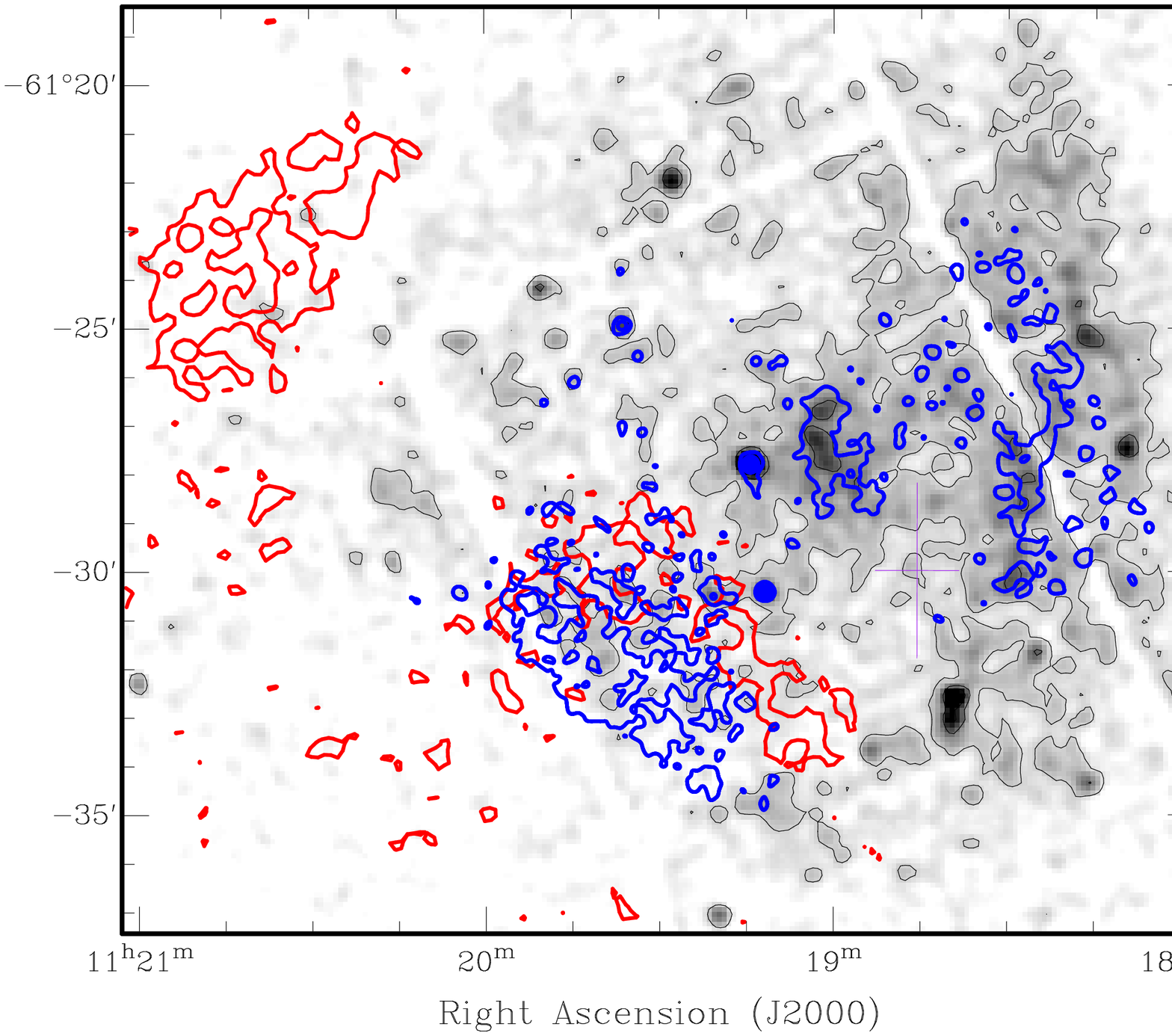}
\vspace{-1cm}
\caption{(a) \textit{Top}:  DSS intensity image of the environment containing SNR G292.2--0.5 (with the pulsar position marked by the purple cross) smoothed using a Gaussian of $\sigma$=5$\arcsec$. DC 292.3--0.4 is shown by the green circle; another dark region appearing to coincide with SNR--East (Region 1) \& PSR--East (Region 2) is shown by the green ellipse. (b) \textit{Bottom}: Contours showing the dark DSS clouds (in red) overlaid on the \textit{XMM-Newton} (in black) and the \textit{Chandra} (in blue) X-ray images. The position of the TeV emission detected with H.E.S.S. and having a centroid between PSR--West and SNR--West (Djannati et al. 2009) is also shown by the purple cross.}
\end{figure*}

\subsection{Origin of non-thermal X-ray emission}
\label{5.4}

As pointed out above and summarized in Table~3, our study confirms the GSH05 result for the presence of hard non-thermal X-ray emission from the SNR. The additional \textit{Chandra} and \textit{XMM-Newton} observations allowed us to fit for two-component models (thermal+non-thermal), resolve the non-thermal emission, and better constrain the spectral properties of both components.

While the emission from the SNR is mostly dominated by thermal emission best described by a non-equilibrium ionization model with $kT$$\geq$1~keV and an ionization timescale of $\sim$10$^{10}$~cm$^{-3}$~s, the emission from most regions (except for SNR--West) required an additional non-thermal component described by a power law model with a hard photon index, $\Gamma$$\sim$1--2. The total non-thermal X-ray luminosity in the studied regions is 6.1$\times$10$^{33}$~$D^2_{8.4}$~ergs~s$^{-1}$, which is $\sim$4\% of the total SNR emission. In the previous section, we discussed the origin of the non-thermal emission from SNR--East as possibly due to the interaction between the SNR and a cloud in the east (see Figure~6).

As to the PSR--West and PSR--East regions\footnote{For PSR--East, the additional PL model was not statistically needed but resulted in a more reasonable temperature for the thermal component (see Section 4.4).}, the  non-thermal emission can be attributed to two scenarios:  (1) the leakage of high-energy particles from the pulsar or its PWN into the SNR, and/or (2) particles accelerated at the supernova shock leading to a synchrotron component in the high-energy end of the X-ray spectrum, as observed in a growing number of non-thermal SNR shells such as G347.3--0.5 (Slane et al. 1999). Both scenarios were discussed in GSH05. Here, we further explore these scenarios in the light of the new data.

In the first scenario, the non-thermal emission is attributed to relativistic particles originating from the PSR or its associated PWN. We favor this scenario due to: a) the location of the selected regions with respect to the pulsar and its associated PWN well inside the SNR shell, and 2) the resemblance between their photon indices. PSR J1119$-$6127 has a photon index of $\Gamma$=1.9$^{+1.1}_{-0.9}$ (using a power-law + blackbody model fit) and its associated compact PWN has a $\Gamma$=1.1$^{+0.9}_{-0.7}$  (Safi-Harb \& Kumar 2008).  An elongated jet-like feature south of the pulsar was also characterized by a similarly hard power-law index ($\Gamma$$\sim$1.4). These values are consistent with the photon index obtained for PSR--East ($\Gamma$=0.9$^{+0.6}_{-0.2}$) and PSR--West ($\Gamma$=1.5$^{+0.5}_{-0.2}$), suggesting a similar particle population. The non-thermal X-ray luminosity from PSR--East and PSR--West amounts to $\sim$3.3$\times$10$^{33}$$D^2_{8.4}$~ergs~s$^{-1}$, which represents only $\sim$0.1\% of the pulsar's spin-down energy, $\dot{E}$.

For the second scenario, the thermal X-ray emission from PSR--East and PSR--West  hints at enhanced metal abundances, suggesting the presence of ejecta heated by the reverse shock. For the non-thermal component, the photon index, $\Gamma$$\sim$0.7--2.0 (Table~3), is however harder than that seen in all non-thermal SNRs (where $\Gamma$$\sim$2--3). As well, the acceleration to high energies in the non-thermal SNR shells is believed to take place at the forward, not the reverse, shock. This, together with the presence of non-thermal emission in the SNR interior, argue against shock acceleration at the supernova shock wave as an origin. Whether acceleration at the reverse shock can explain at least part of the X-ray emission remains an open question to be explored with further observations and modeling.

In summary, we favour the model for non-thermal emission arising, at least partly, from relativistic particles injected by the pulsar or its PWN. This is further corroborated by the detection of TeV emission offset from the pulsar with H.E.S.S. (Djannati et al. 2009),  with its peak located between PSR--West and SNR--West as shown by the purple cross in Figure~6 (b).  Although the origin of TeV emission is not yet clearly identified, Djannati et al. (2009) suggest that the emission could be attributed to an offset PWN, where the nebula has been displaced after being crushed by an asymmetric reverse shock caused by the presence of the dark cloud in the north-east. An origin for the gamma-ray emission in cosmic ray acceleration at a supernova shock requires a higher ambient density ($\sim$1~cm$^{-3}$, Djannati et al. 2009) which is ruled out by our spectral study. We can not however rule out a higher density medium in the regions that we could not study either due to lack of coverage by the X-ray instruments or due to low-count spectra, especially in the south-west where the TeV emission peaks. The faint and hot gas seen however from the southern region (covered only by \textit{XMM-Newton}, see section 4.5) suggests expansion in a more tenuous medium.

\subsection{Progenitor mass of SNR G292.2--0.5}
\label{5.5}

One of the primary objectives of this paper is to estimate the progenitor mass of SNR G292.2--0.5. This goal is motivated by studying the connection between high-magnetic field pulsars and magnetars through studying their environments/associated SNRs. In the following, we discuss the implication on the progenitor mass for SNR G292.2$-$0.5 from our X-ray spectroscopy, and discuss previously suggested models for the progenitor mass of this SNR and other highly magnetized neutron stars.

X-ray spectroscopy has proven to be a powerful tool to infer the mass of progenitor stars for young SNRs dominated by shock-heated ejecta.
By comparing the abundances of heavy elements inferred from fitting X-ray spectra to the nucleosynthesis yields inferred from
core-collapse models such as those of Woosley \& Weaver 1995 (hereafter WW95) and Thielemann et al. 1996 (hereafter TNH96),
the mass of the progenitor star can be determined. WW95 provides nucleosynthesis yields for core-collapse supernovae of varying masses ranging from 13--40$M_{\sun}$ whereas TNH96 provides the yields specifically for the 13, 15, 20 and 25$M_{\sun}$ progenitor stars. However, the yields obtained from these two models for a particular progenitor star differ due to the different assumptions used in the processes such as convection, explosion energy and mechanism, mass-loss, and initial metallicity (Hoffman et al. 1999).  The TNH96 models produce more Ne and Mg, and less Fe in the more massive ($\ge$20--25 $M_{\sun}$) stars than the WW95 models, while they agree fairly well with each other for the lower mass stars ($\le$20$M_{\sun}$). Also, WW95 specify models A, B, and C for stars of progenitor masses $\ge$30$M_{\sun}$. The difference between these models is that the final kinetic energies at infinity ($KE_{\infty}$) of the ejecta for models B and C are enhanced by a factor of $\sim$1.5 and $\sim$2, respectively, with respect to model A ($KE_{\infty}$$\sim$1.2$\times$10$^{51}$~ergs). Keeping in mind the above differences, the inference on the progenitor masses may likely be different depending on the models used. In our studies, we focus on the WW95 nucleosynthesis models which provide a wider range of progenitor masses and finer mass steps.

The nucleosynthesis models mentioned above show that the mass or relative abundance ratio [X/Si]/[X/Si]$_{\sun}$ of ejected mass of any element X with respect to Silicon (Si) varies significantly as a function of the progenitor mass. Under the assumption of the detection of ejecta in G292.2--0.5 (see Section 5.1 and Table~3), we subsequently estimate the abundance ratios for the reverse-shocked regions. In Figure 7, we plot the average values of these ratios and their root mean square (RMS) scatter (shown in red). Also, over-plotted are the predicted WW95 relative abundances (estimated from the derived production factors after radioactive decay) for different progenitors of masses 15, 18, 20, 25, 30, and 35 $M_{\sun}$. We find that the average abundance ratios fit the expectation of a 30$M_{\sun}$ progenitor star within error. We also compare our values with the TNH96 nucleosynthesis model yields for 13, 15, 20 and 25$M_{\sun}$ progenitors and find that our best-fit values are consistent with the 25 solar-mass model.

\begin{figure*}
\center
\includegraphics[width=0.8\textwidth]{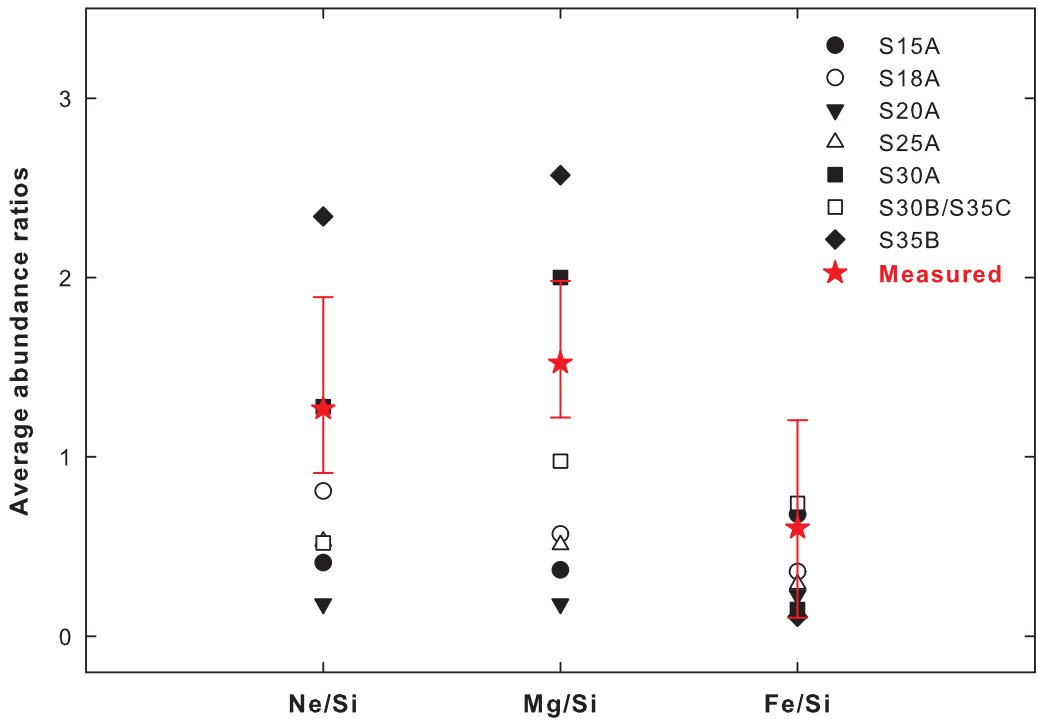}
\caption{Best-fit abundances of Ne, Mg and Fe, relative to Si relative to solar (plotted in red). Also, over plotted are the predicted relative abundances [X/Si]/[X/Si]$_{\sun}$ from the core-collapse nucleosynthesis models of WW95. We combine the nucleosynthesis yields of progenitors of masses 30B and 35C as their elemental abundance ratios with respect to Si differ very slightly. See text for details.}
\end{figure*}

We conclude that both the WW95 and TNH96 models predict a high-mass ($\sim$25--30$M_{\sun}$) progenitor star for SNR G292.2-0.5 and that our best fit abundance values rule out lower mass progenitors. We caution the reader however that this conclusion is based on three data points and on the fact that the abundances are only slightly enhanced. More sensitive observations are needed to confirm the presence of ejecta and better constrain the metal abundances, including O and Fe which are important diagnostics for the progenitor mass estimate, but were at the limit of our detection. 

Chevalier (2005) suggests that PSR J1119--6127/SNR G292.2--0.5 may belong to either the SN IIP or Ib/c categories. Theoretical models suggest that SNe IIP come from stars of mass $\sim$10--25$M_{\sun}$ (Heger et al. 2003) with relatively low-mass rates and are expected to explode as red supergiants (RSGs) with most of their H envelope present, giving rise to the extended plateau emission. During the supernova explosion resulting in a SN IIP, the massive H envelope is able to decelerate the core material and the reverse shock wave carries H back toward the center. Also, due to their low mass loss, the RSG winds extend only to a small distance of $\le$1~pc from the progenitor and is surrounded by a low-density wind bubble created during the main-sequence phase (Chevalier 2005). On the other hand, SNe Ib/c result from Wolf-Rayet (W-R) stars of mass $\ge$30--35$M_{\sun}$ (Heger et al. 2003), formed after a RSG phase or a brief luminous blue variable phase during which they have shed most of their outer H envelopes. W-R stars are characterized by high mass-loss rates ($\dot{M}$$\sim$10$^{-5}$$M_{\sun}$~yr$^{-1}$) and high-velocity stellar winds ($v_w$$\sim$1000~km~s$^{-1}$). The fast moving W-R winds can sweep up the circumstellar and slow-velocity wind materials, clearing out a low-density wind bubble around the star which can extend to $>$10~pc over a duration of 2$\times$10$^{5}$~yr (Chevalier 2005). The inferred low ambient density (Table~4), low SNR X-ray luminosity, and the high progenitor mass inferred above from our X-ray spectroscopic study, all point to a W-R progenitor star, thus ruling out the SN IIP progenitor.  

We now briefly discuss the progenitor mass studies done on magnetars and other high-magnetic field pulsars. It had been previously suggested by a few authors that magnetars descend from massive progenitors. For example, the progenitor masses of the soft gamma repeaters (SGRs) 1806--20 and CXOU J164710.2-455216 associated with the clusters Cl 1806--20 and Westerlund 1 (Wd 1) were determined as 48$^{+20}_{-8}$$M_{\sun}$ and 40$\pm$5$M_{\sun}$, respectively (Figer et al. 2005; Bibby et al. 2008; Muno et al. 2006). The progenitor mass of the anomalous X-ray pulsar (AXP) 1E 1048.1--5937  was estimated as 30--40$M_{\sun}$ from the study of the expanding HI shell around it, which was interpreted as a stellar wind bubble blown by the progenitor (Gaensler et al. 2005). However, a recent study by Davis et al. (2009) suggests a lower progenitor mass of 17$\pm$2$M_{\sun}$ for the magnetar SGR 1900+14 inferred from the infrared studies of its association with stellar clusters, challenging the massive progenitor predictions. A recent study of the SNR Kes~73 associated with the AXP 1E 1841--045 using \textit{Chandra} and \textit{XMM-Newton} also suggests a progenitor mass $\sim$25--30$M_{\sun}$ (Kumar et al. 2010). Out of the seven high-magnetic field pulsars discovered so far, only two of them (PSRs J1119--6127 and J1846--0258) are so far associated with SNRs.  PSR~J1846--0258 associated with the SNR Kes~75 recently displayed a magnetar-like behaviour (Gavriil et al. 2008; Kumar \& Safi-Harb 2008).  The  \textit{Chandra} and \textit{Spitzer} study of SNR Kes 75  implied, like for J1119--6127, a W-R progenitor star for the remnant (Morton et al. 2007; Su et al. 2009).  

In summary, while most current studies seem to be pointing to highly magnetized neutron stars having massive progenitors ($\geq$25$M_{\sun}$), further studies in X-rays and other wavelengths are needed to address this interesting question.

\section{Conclusions}
\label{6}

We have presented the first detailed imaging and spatially resolved X-ray spectroscopic study of  the Galactic SNR G292.2--0.5, associated with the high-magnetic field pulsar, PSR J1119$-$6127, combining all the available \textit{Chandra} and \textit{XMM-Newton} observations. The regions selected for spectroscopy are shown in Figure 2.  We summarize here our findings:
\begin{itemize}
\item{The high-resolution images show diffuse emission across the SNR, with brighter and softer emission in the west. While the SNR has been classified as a shell-type remnant in the radio, the X-ray morphology suggests a composite-type SNR, with both thermal and non-thermal X-ray emission from the interior regions. An elongated and bright limb is particularly visible in the west in both the \textit{Chandra} and \textit{XMM-Newton} images, which partially coincides with the SNR shell seen in the radio (see Figure~1).}

 \item{The diffuse X-ray emission from the SNR is dominated by thermal emission from hot plasma, best described by an absorbed plane-parallel shock, non-equilibrium ionization model (VPSHOCK in XSPEC) with a  temperature $kT$$\geq$1~keV (ranging from  $\sim$1.3 keV in SNR--West to $\sim$2.3 keV in SNR--East) and with an ionization timescale of $\sim$10$^{10}$ cm$^{-3}$~s (ranging from $\sim$6$\times$10$^{9}$ cm$^{-3}$~s in the interior to $\sim$3$\times$10$^{10}$ cm$^{-3}$~s in SNR--West) indicating that the plasma has not yet reached ionization equilibrium -- a result that is consistent with a young SNR expanding in a very low-density medium. The thermal fits indicate the presence of slightly enhanced abundances from Ne, Mg, Si, hinting for the first time at the presence of ejecta heated by the reverse shock. The metal abundances inferred for SNR--West are consistent with solar or sub-solar values. This, together with the higher ionization timescale and the morphology of this region, suggest that its emission arises from the supernova blast wave or shocked circumstellar matter.}
 
 \item{The eastern side of the remnant is fainter than the western side and is characterized by harder X-ray emission ($kT$=2.3$^{+2.9}_{-0.5}$ keV) and a higher column density ($N_H$=1.8$^{+0.2}_{-0.4}$$\times$10$^{22}$ cm$^{-2}$, $\sim$8$\times$10$^{21}$ higher than in SNR--West). An investigation of the eastern field with optical DSS data suggests the presence of two dark clouds: one to the north-east coinciding with DC 292.3$-$0.4, a dark cloud catalogued by Hartley et al. (1986) and discussed by P01,  and another, previously uncatalogued, dark cloud to the south-east overlapping the emission from SNR-East and PSR-East. These spectral and imaging properties are consistent with the presence of an intervening cloud to the east that would explain the lack of soft X-ray emission from the eastern side of the remnant.}
 
 \item{An additional non-thermal component, described by a power-law model with a hard photon index ($\Gamma$$\sim$0.7--2.0) and representing $\sim$4\% of the total flux, is needed for the regions in the east (SNR--East and PSR--East) and in the western region close to the pulsar (PSR--West). The emission from SNR--East could be attributed to the interaction with the nearby cloud, although the photon index appears unusually hard and so far there is no evidence of TeV or OH maser emission from this region. The emission from PSR--East and PSR--West can be attributed to leakage of relativistic particles from the pulsar or its associated PWN. The detection of TeV emission with H.E.S.S. from the south-western region (between PSR-West and SNR-West) supports this conclusion, although a further investigation of this region is needed.}

\item{Using an SNR diameter of $\sim$17$\arcmin$.5 and a distance of 8.4~kpc, we derive an SNR age ranging between  4.2~$D_{8.4}$ kyr (free expansion phase, assuming an expansion velocity of 5000 km~s$^{-1}$) and  7.1~$D_{8.4}$~kyr (Sedov phase). For the Sedov phase estimate, we assume that the SNR-West region represents emission from the blast wave shocking the interstellar or circumstellar medium. This SNR age range is a factor of a few higher than the pulsar's estimated age upper limit of 1.9~kyr. We discuss the discrepancy between the SNR and pulsar age estimates and conclude that this may be attributed to a variable braking index for the pulsar, which has also shown recently some unusual timing characteristics at radio wavelengths. Requiring the SNR's age to be $\leq$1.9~kyr implies  an unusually high expansion velocity $\gtrsim$11200~$D_{8.4}$~km~s$^{-1}$ throughout its life, which is at odds with the observations (especially given the asymmetry of the remnant and since the SNR appears to be transitioning from the ejecta-dominated, free expansion phase, to the Sedov phase). The Sedov phase gives a lower limit on the  shock velocity of $\sim$1.1$\times$10$^3$~km~s$^{-1}$ and an explosion energy of $\geq$6$\times$10$^{50}$~$f^{-1/2}D^{5/2}_{8.4}$ ergs, under the assumption of an explosion in a uniform ISM. The total X-ray luminosity from the studied SNR regions is $L_X$ (0.5--10 keV) $\geq$1.7$\times$10$^{35}$ $D^2_{8.4}$~ergs~s$^{-1}$.}

 \item{Using the enhanced metal abundances inferred from our fits and comparing to core-collapse nucleosynthesis model yields, we infer a progenitor mass of $\sim$30$M_{\sun}$, a result that supports a SN type Ib/c origin and other studies suggesting very massive progenitors for the highly magnetized neutron stars.}
 
 \item{We have studied other (fainter)  regions in the SNR in the north and south covered only by the \textit{XMM-Newton} observation. Our fits
indicate also the presence of hot plasma that has not reached ionization equilibrium, possibly combined with non-thermal emission, however the statistics did not allow us to constrain the models. Deeper and more sensitive observations that cover the entire SNR and the radio shell are needed to better constrain the X-ray emission  throughout the remnant and confirm the metal abundances in the SNR.}
 
 \item{Targeted CO, and other wavelength, studies of the environment of G292.2$-$0.5 are needed to confirm or rule out any interaction with a nearby molecular cloud, and to explain the TeV emission detected with H.E.S.S. in the south-west.}
 
\end{itemize}

\acknowledgements
This research made use of NASA's Astrophysics Data System and of NASA's HEASARC maintained at the Goddard Space Flight Center. S. Safi-Harb acknowledges support by a Discovery Grant from the Natural Sciences and Engineering Research Council, the Canada Research Chairs program,  the Canada Foundation for Innovation, and the Canadian Space Agency.  We thank Jayanne English for her help with overlaying the radio contours in Figure~1 and for insightful discussions on the association with the cloud. We are grateful to the referee for a thorough reading of the paper and useful comments that helped improve the paper.

\end{document}